\documentclass{jfm}

\usepackage{graphicx}
 \usepackage{newtxtext}
 \usepackage{newtxmath}
 \usepackage{amsmath}
 \usepackage{upgreek}
 \usepackage{xcolor}
 \usepackage{mathrsfs}

 \usepackage{natbib}
  \usepackage{hyperref}
  \hypersetup{
      colorlinks = true,
      urlcolor   = blue,
      citecolor  = black,
  }

\newcommand{\RomanNumeralCaps}[1]
\linenumbers

\title{Non-adiabatic modulation of premixed-flame thermoacoustic frequencies in slender tubes}

\shorttitle{Non-adiabatic thermoacoustic frequencies}
\shortauthor{E. Flores-Montoya and others}

\author{Enrique Flores-Montoya\aff{1},
Victor Muntean \aff{1},
Mario S\'anchez-Sanz \aff{2}
\and Daniel Mart\'inez-Ruiz \aff{1}
  \corresp{\email{}}}

\affiliation{\aff{1} ETSIAE, Universidad Polit\'ecnica de Madrid, 28040, Madrid, Spain
\aff{2} Universidad Carlos III de Madrid, Legan\'es, Spain}

\begin{document}

\maketitle


\begin{abstract}
This paper presents an experimental study of the influence of heat losses on the onset of thermoacoustic instabilities in methane-air premixed flames propagating in a horizontal tube of small diameter, $D = 10$~mm. Flames are ignited at the open end of the tube and propagate towards the closed end undergoing strong oscillations of different features due to the interaction with acoustic waves. Different regimes are reproduced in these experiments, namely the well-known primary and secondary thermoacoustic instabilities, which are strongly dependent on mixture equivalence ratio, fuel properties and geometric constraints of the combustion chamber. The frequency of oscillation and its axial location are controlled through the tube length $L$ and the intensity of heat losses. These parameters are respectively modified in the experiments by a moveable piston and a circulating thermal bath of water prescribing temperature conditions.

Main experimental observations show that classical one-dimensional predictions of the oscillation frequency do not accurately describe the phenomena under non-adiabatic real scenarios. Besides the experimental measurements, a quasi-one-dimensional heat transfer analysis of the burnt gases downstream of the flame is provided. {\color{black}This study introduced the effect of heat losses at the walls of the tube on the interplay between the acoustic field and the reaction sheet. As a result, this analysis provides an improved description of the interaction and accurately predicts the excited flame-oscillation harmonics through the eigenvalues of the non-adiabatic acoustics model}. Unlike the original one-dimensional adiabatic analysis, {\color{black} the comparison between the flame oscillation frequency provided by the non-adiabatic extended theory and the frequencies measured in our experiments is in great agreement in the whole range of temperatures considered}. This confirms the importance of heat losses in the modulation of the instabilities and the transition between the different flame oscillation regimes.

\end{abstract}


\section{Introduction}\label{sec:intro}

Premixed flame propagation is a rich problem that has been fruitfully addressed for many years by the scientific community. Numerous technological applications rely on the fundamental understanding of the physical processes that take part in the aerothermochemistry of reacting fronts. In particular, the knowledge on the coupling between flames and acoustic waves is essential to an adequate design and operation of most burners and combustion chambers in favor of safety and efficiency. \textcolor{black}{Thermoacoustic coupling in tubes has been a field of intense study since the original reports by \citet{sondhauss1850ueber}, dealing with external heating of pipes with a closed end, and by \citet{rijke1859} in heated tubes with two open ends. Regardless of the heat release source, this interaction was first explained by \citet{rayleigh1878explanation} through his thermoacoustic growth criterion: heat promotes acoustic waves if released at the compression stage or if extracted during rarefaction. This behavior was detected early on burners} by \citet{mallard1881vitesses} and later addressed experimentally in \citep{mason1920v,coward1937}, where the marked behavior of oscillating flames under smooth and violent regimes was directly related to acoustic coupling.

First theoretical models to explain thermo-acoustic instabilities \textcolor{black}{with flames} appeared in the context of project SQUID after World War II \citep{markstein1950squid}, which proposed the so-called parametric instability. The latter is defined as the coupling driven by an imposed oscillating-pressure flow interacting with a flame \citep{markstein1953instability,markstein1955stability}. These analyses lead to Mathieu's equation, that was able to provide a description of the flame response to the acoustic field given the amplitude and frequency of the oscillatory velocity and the wavelength of the perturbed reacting front. The stability problem was later revisited by other authors \citep{searby1991parametric,aldredge2004experimental, yanez2015flame} under the assumption of a fixed acoustic field. However, the extended theoretical analyses of the flame response to self-excited parametric acoustic oscillations have only been offered recently \citep{wu2009flame,assier2014linear}. \textcolor{black}{Other oscillatory regimes have been found in stabilized premixed flames in tubes \citep{richecoeur2005experimental}, related to extinction-reignition cycles for diameters of the tube comparable to the quenching distance and strongly influenced by the temperature and thermal properties of the tube material \citep{evans2009operational}. }

Markstein also reported the arising acoustic harmonics, their position along the tube upon flame coupling, and smooth-to-strong excitation events found for hydrogen, methane and higher hydrocarbons at either rich or lean mixtures with air \citep{markstein1951interaction}. This change in behavior was later defined as the development of primary and secondary acoustic instabilities, reported in propane flames traveling downwards towards a closed end of a tube with the ignition end open to the atmosphere \citep{searby1992acoustic}. Primary acoustic instabilities are characterized by a small pulsation amplitude and smooth variations of the flame surface. However, secondary acoustic instabilities involve large-amplitude oscillations and pressure variations at least one order of magnitude greater than in the former case. In addition, the flame front is typically corrugated and disordered, producing a large increase in burning area and overall reaction rate. However, accurate modelling and explanations of the mechanisms controlling this kind of transition remain under discussion. Nevertheless, similar transitions were also found for downward propagating methane flames in a periodic Taylor-Couette burner \citep{aldredge2004experimental}, where rich flames showed a more stable behavior than lean flames. Experiments in narrow tubes open at both ends show equivalent trends for propane flames, which propagate with large- and small-amplitude oscillations depending on the stoichiometry of the mixture \citep{connelly2014,yang2015oscillating}. Furthermore, experiments in Hele-Shaw burners have reproduced the transition from primary to secondary instabilities for different fuels, which seem to take place for richer mixtures in the case of propane and DME, but for leaner mixtures in hydrogen and methane premixed flames  \citep{veiga2019experimental,veiga2020thermoacoustic,martinez2019premixed}, ruling out simplified explanations based on flame temperature and flame propagation velocity criteria alone.

Numerical studies have also provided detailed information on the onset and transition between the two different oscillatory regimes for a flame propagating towards the closed end of narrow domains. Acoustic oscillations have been shown to produce an acceleration field at the flame front leading to intense Rayleigh-Taylor instabilities that wrinkle the flame-front in sufficiently wide regions \citep{petchenko2006violent}. However, small-amplitude flame oscillations are preserved in domains where the channel height $h$ and the flame thickness $\delta_T$ hold a moderate scale relation $h/\delta_T \simeq 10$ \citep{petchenko2007flame}. This effect, directly related to viscous damping, has been reported for thermoacoustic instabilities of lean hydrogen-air mixtures in Hele-Shaw cells \citep{veiga2020thermoacoustic}, where thinner channels drastically reduce the acoustic coupling. In turn, small-diameter tubes show the same behavior in suppressing and extinguishing premixed flames under parametric instabilities \citep{dubey2021acoustic}. 

In addition, simulations of axisymmetric flames propagating upwards exhibit the same oscillating quasi-stationary response to acoustic perturbations \textcolor{black}{as in the experiments}, until wrinkles arise in the reacting front \citep{higuera2019acoustic}. Other numerical studies conclude that symmetric and non-symmetric flames propagating in tubes in absence of gravity present different gain to acoustic wave perturbations and varying onset of the instability with non-negligible dependence on the channel length, thickness, and on the chemical kinetics modelling \citep{jimenez2021flame}. 

All things considered, secondary acoustic instabilities seem to \textcolor{black}{require the previous development of} primary acoustic oscillations to  \textcolor{black}{take place}. Therefore, numerous works have proposed fundamental explanations to the origin of the finite-amplitude initial coupling \citep{clavin1990one,clanet1999primary}, with two mechanisms considered to this end. First, the modification of the internal structure of the flame front due to the sensitivity of the chemical kinetics to temperature variations. This mechanism was addressed through a correlation between the product $\beta M$, being $\beta$ and $M$ the Zel'dovich and the Mach number respectively, and the acoustic pressure in \citet{yoon2016onset}. This suggested a non-negligible sensitivity of the reaction rate to acoustic pressure variations through temperature. Nevertheless, it was theorized in \citet{clavin1990one} that this amplification due to acoustic temperature fluctuations was relevant only in marginal cases. Second, the geometry variation of the flame related to the sensitiveness of the curved flame front to the presence of an acceleration as given by the acoustic field. A theoretical analysis of this velocity coupling in the limit of weakly-wrinkled flames was carried out in \citet{pelce1992vibratory}, with predicted growth rates two orders of magnitude greater than the pressure-coupling feedback.
\textcolor{black}{In spite of} this, both analyses provide a common first-order prediction of the coupling frequencies given by the outer flow field, independent of the considered transfer mechanism at the flame inner region.

These theoretical studies on the initiation mechanisms have produced detailed predictions of the modification of the flame structure, transfer function and oscillating frequency along the tube. Although very useful in the comparison of thermoacoustic simulations with adiabatic boundaries \citep{jimenez2021flame}, the oscillation frequency is not accurately described when heat losses are not negligible. \textcolor{black}{For this reason, introduction of conductive heat-transfer processes through the walls of the system is mandatory to represent real scenarios. For instance, Rijke-type devices, related to generic heat sources in open tubes \citep{raun1993review}, have been recently proposed as a benchmark to control acoustic oscillations via heat transfer from secondary sources \citep{jamieson2017experimental}. An adequate comprehension of the non-adiabatic effects is thus needed to progress in the description and understanding of premixed-flame thermoacoustic instabilities.} The outer-flow acoustic model of \citet{clavin1990one} and \citet{pelce1992vibratory} is here addressed to propose an extension to non-adiabatic thermal effects that yields adequate theoretical results for the flame oscillation frequency, which is in close agreement with the experiments.

Most canonical experimental studies found in the literature focused their attention on flames propagating in vertical tubes under the effect of gravity. In this paper, a non-adiabatic extension to the one-dimensional coupling frequency model is offered to accurately predict the oscillations of premixed flames in \textcolor{black}{very} slender configurations. Later, a horizontal tube of inner diameter $D$, such that the Froude number is ${\rm Fr}\simeq 1$ and transverse buoyancy effects are non-dominant, is used together with a circulating thermal bath to capture the controlling effects of the oscillating frequency. This \textcolor{black}{frequency} value, which sustains first-order modifications, plays a major role in the initiation of thermoacoustic instabilities. Furthermore, the selection of a moderate diameter aims to avoid complex aerodynamic structures that can take part in the fast wrinkling of the flame by setting the experimental cases in the nearly-viscous limit. Finally, a \textcolor{black}{moveable} piston allows modifications to the tube length $L$, \textcolor{black}{measured from the piston to the open end of the tube}, which controls the main acoustic frequency. \textcolor{black}{The latter also depends on the varying position of the flame $r$ along the tube, providing the evolution of the oscillation frequency during flame propagation.} The experimental setup is, thus, envisioned to validate the extended model proposed here, to simplify the analysis and to facilitate the understanding of thermoacoustic coupling initiation mechanisms.

Additional results of this study include the modification of primary-to-secondary instability transition due to mixture temperature variation. In particular, slightly preheated mixtures are increasingly driven to stabilization in primary acoustic oscillations by thermal suppression of transition to secondary instability regimes. Experimental evidence of this feature is offered in the final section of the manuscript.

\section{Background and theoretical modeling}

\subsection{Primary acoustic instabilities}

The existing theoretical predictions of the coupling and amplification of primary thermoacoustic modes follow the seminal works of \citet{clavin1990one} and \citet{pelce1992vibratory}, which provide a method to compute the acoustic frequencies of a one-dimensional flame propagating from the open to the closed end of a tube. From the perspective of the fluid dynamics acoustic problem, the flame front is seen as a discontinuity surface separating the unburnt and the burnt gases when the dimensionless activation energy is a large parameter, be the Zel'dovich number $\beta \gg 1$. 

\begin{figure}
    \centering
	\includegraphics[width=0.7\textwidth]{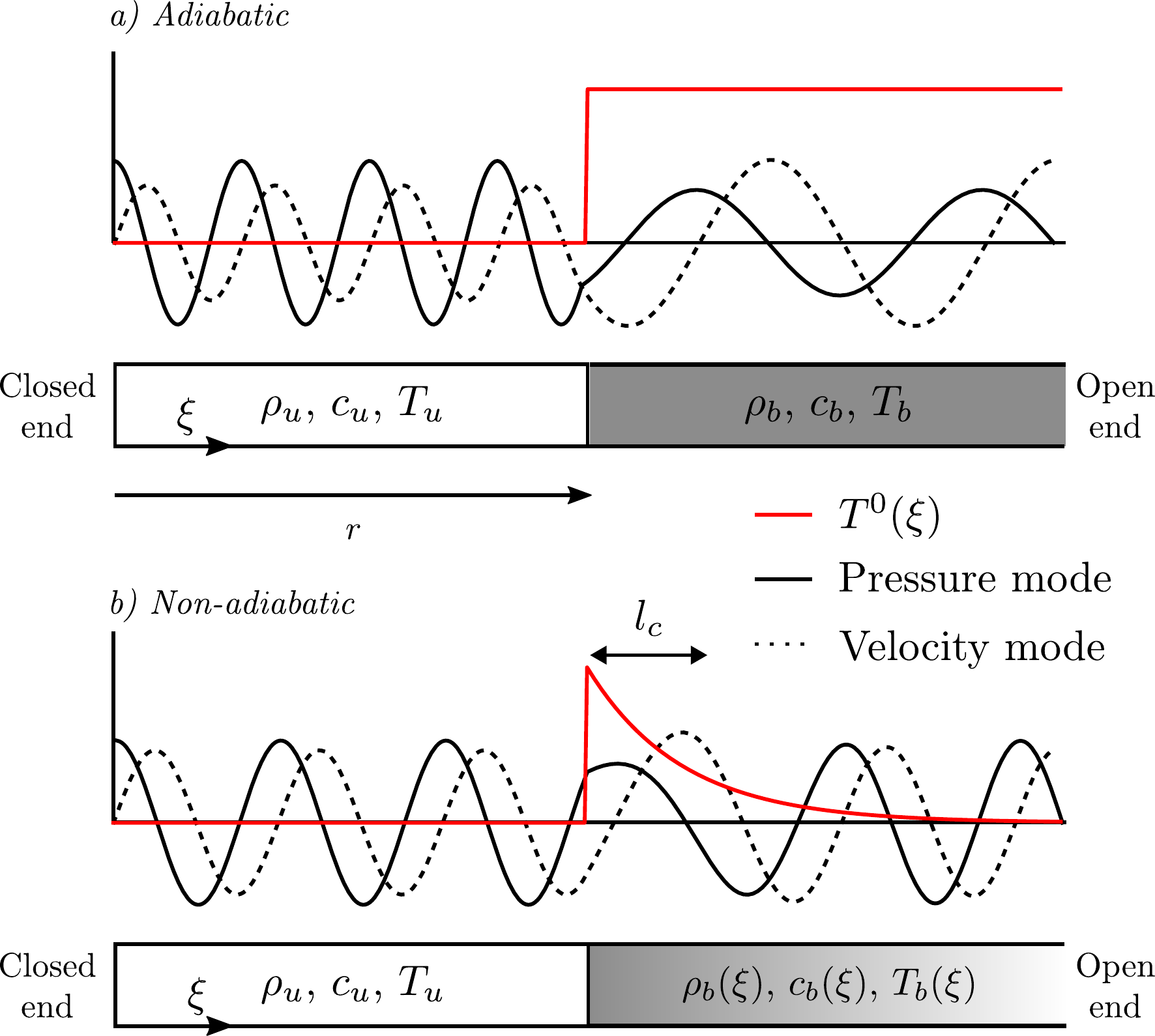}
	\caption{ Sketch of adiabatic and non-adiabatic conditions of flame propagation in a tube.}
	\label{fig:sketch_acoustics.pdf}
\end{figure}

A sketch of the gas properties distribution applicable to this model is presented in {Fig.~\ref{fig:sketch_acoustics.pdf}a} along with the associated pressure and velocity modes. The dimensionless axial coordinate $\xi = x/L$ measures the distance to the closed end of the tube and the flame sheet is located at $\xi = r$. Therefore, in region $0<\xi<r$ the properties of the gas are those of the quiescent unburnt mixture, be density $\rho_u$ and temperature $T_u$, and the region with $r<\xi<1$ exhibits burnt gas properties, $\rho_b$ and $T_b$, where $T_b = \epsilon T_u$ is the adiabatic flame temperature. Due to the dependence of the speed of sound on fluid temperature, $c=\sqrt{\gamma \mathcal{R}_gT}$, acoustic waves propagate at different velocities in the unburnt, $c_u$, and burnt, $c_b$, gas sections. For this reason, as the flame propagates along the tube at velocity $S_u\ll c_u$, the length of the region occupied by combustion products increases, and a variation of the natural frequency of the tube is obtained.

The linearized analysis is performed over the perturbation of base-flow variables denoted by superscript $0$, via small acoustic variations denoted by superscript $1$, {\it e.g.}  $u=u^0(\xi) + u^1(\tau,\xi)$, and is applied at both sides of the flame. The dimensionless momentum and mass governing equations for the perturbations read,
\begin{equation}\label{eq:ac_mom}
    \frac{\rho^0}{\rho_u}\frac{\partial \hat{u}}{\partial \tau} = - \frac{\partial \pi}{\partial \xi}, \qquad
    \frac{\partial \pi}{\partial \tau} = - \frac{\rho^0}{\rho_u}\left(\frac{c^0}{c_u}\right)^2 \frac{\partial \hat{u}}{\partial \xi},
\end{equation}
where $\hat{u}=u^1/S_u$, $\pi = p^1 /(\rho_u S_u c_u)$ and $\tau = tc_u/L$. Combination of both expressions allows to write the problem in the wave equation form,
\begin{equation}
    \frac{\partial^2 \hat{u}}{\partial \tau^2} =  \Delta(\xi)\frac{\partial^2 \hat{u}}{\partial \xi^2}, \qquad
    \frac{\partial^2 \pi}{\partial \tau^2} =  \frac{\partial}{\partial \xi}\left[\Delta(\xi)\frac{\partial \pi}{\partial \xi}\right],
    \label{eq:wave_u}
\end{equation}
where $\rho^0T^0 = p^0/\mathcal{R}_g = \mathrm{const}$, in the framework of the acoustic analysis. Moreover, for the sake of generality and as an extension to \citep{clavin1990one,pelce1992vibratory}, the temperature ratio $\Delta(\xi) = T^0/T_u$ is here allowed to follow a non-uniform distribution as sketched in Fig.~\ref{fig:sketch_acoustics.pdf}b. An equation for the natural frequencies of the tube with two regions of different properties can be derived by solving eqs.~\eqref{eq:wave_u} with the appropriate boundary conditions at both ends of the tube for acoustic velocity and pressure perturbations, 
\begin{equation}
    \hat{u}(\xi=0) =\pi(\xi=1) =0.
    \label{eq:bcs}
\end{equation}

In addition, jump conditions across the flame sheet must be provided for adequate problem closure. Typical low-Mach flame propagation produces pressure variations across the discontinuity of order $\sim\rho_u S_u^2$, much smaller than acoustic pressure variations of order $\sim\rho_u S_u c$, which yields 
\begin{equation}
    \pi(\xi=r^-) = \pi(\xi=r^+).
    \label{eq:jcp}
\end{equation} 
Furthermore, first-order solution assumes the flame front to behave as a passive interface, such that \begin{equation}
    \hat{u}(\xi=r^-) = \hat{u}(\xi=r^+) = (L/{S_u})(dr/dt).
    \label{eq:jcu}
\end{equation}

The solution to the one-dimensional acoustic perturbation problem can be analytically derived for constant temperature of the burnt gases $\Delta = T_b/T_u = \epsilon$, and $\Delta = 1$ in the unburnt mixture 
\begin{eqnarray}
    &\hat{u}_u = e^{i\Omega\tau}\left[A_u e^{i\Omega\xi} + B_u e^{-i\Omega\xi}\right], 
   & \pi_u = -e^{i\Omega\tau}\left[A_u e^{i\Omega\xi} - B_u e^{-i\Omega\xi}\right],  \label{eq:u_pi_burnt_Tconst} \\
    &\hat{u}_b = e^{i\Omega\tau}\left[A_b e^{i\Omega\xi/{\sqrt{\epsilon}}} + B_b e^{-i\Omega\xi/{\sqrt{\epsilon}}}\right], 
   & \pi_b = -\frac{e^{i\Omega\tau}}{\sqrt{\epsilon}}\left[A_b e^{i {\Omega}\xi/{\sqrt{\epsilon}}} - B_b e^{-i{\Omega}\xi/{\sqrt{\epsilon}}}\right]. \nonumber
\end{eqnarray}
The equation for the dimensionless acoustic frequencies $\Omega = \omega t_a$, where $t_a = L/c_u$ is the acoustic time, is prescribed after imposing boundary \eqref{eq:bcs} and jump conditions \eqref{eq:jcp}-\eqref{eq:jcu} to the linear system \eqref{eq:u_pi_burnt_Tconst} for the non-trivial value of constants $A_u$, $B_u$, $A_b$ and $B_b$. Namely, the adiabatic problem yields the analytic expression for the acoustic frequency eigenvalues,
\begin{equation}\label{eq:Clavin_freq}
    \frac{1}{\sqrt{\epsilon}}\tan(\Omega r)\tan\left(\frac{\Omega}{\sqrt{\epsilon}}(1-r)\right) = 1.
\end{equation}

Unfortunately, this simplified analysis does not adequately predict the evolution of the natural frequencies of the tube with the \textcolor{black}{varying} flame position $r$ in the experimental setup, as will be shown below. Further efforts have introduced acoustic and radiative losses as main deviations from the theoretical predictions of stability \citep{clavin1990one,schuller2020dynamics}. However, additional thermal effects in non-adiabatic long and thin tubes at constant wall temperature may substantially modify the burnt-region isothermal picture presented above, causing major variations in the acoustic coupling of the system upon introduction of non-constant distributions $\Delta(\xi)$ in eqs.~\eqref{eq:wave_u}.

\subsection{Heat-loss estimates, cooling length}

Heat losses influence thermoacoustic instabilities of flames in two significant ways. On the one hand, they affect the inner reaction region by extracting part of the energy of the combustion process and reducing the local temperature. In sufficiently thin tubes, heat losses can even disable self-sustained combustion processes, restraining the propagation of premixed laminar flames \textcolor{black}{and even induce extinction-reignition oscillations \citep{richecoeur2005experimental}}. Conversely, in wide tubes where the surface to volume ratio is small enough, heat-loss effects are confined to a very thin region close to the tube wall. 
The present study focuses in intermediate scenarios, moderate tube diameters that prevent the growth of large-scale dynamic instabilities of the flow, and which characteristically sit far from negligible heat losses in the averaged section. In this context, heat losses affect the flame inducing the curvature of the front and being responsible for deviations from simplified one-dimensional reactions. On the other hand and most importantly, the decrease of burnt gas temperature along the tube due to conductive heat losses produces a first-order modification in the velocity at which acoustic waves propagate in this media, changing the coupling frequency. In particular, the variation of the speed of sound with the distance to the flame front can significantly modify the eigenvalues of the acoustic problem as depicted in Fig.~\ref{fig:sketch_acoustics.pdf}. Therefore, the temperature decay behind the flame requires the characterization underneath.

The infinitely-thin flame approximation that propagates with velocity $S_u$ remains valid here for the estimate of the stream-wise cooling distance $l_c$, \textcolor{black}{in which the average burnt gas temperature decreases to reach values of the order of the ambient temperature $T_u$}. On the moving frame of reference, burnt gas temperature equals the adiabatic flame temperature immediately downstream from the flame, such that ${T_b}/{T_u} ={\rho_u}/{\rho_b} = {S_b}/{S_u} = \epsilon$ due to mass conservation and the gas equation of state. However, heat losses through the wall cause the temperature of the products to decrease with distance measured from the flame front. The region near the wall affected by conduction grows accordingly until, at some point, it is comparable to the tube radius $R$, defining the cooling distance from the flame $l_c$.
Although exact calculation of temperature profiles is a complex task, some assumptions can be made to obtain a valuable approximation. 

The lack of chemical reaction in the burnt region enables a straight-forward balance between the convective transport and the radial conduction terms in the energy conservation equation, as the longitudinal conduction term is neglected in slender configurations  \citep{hicks1947one}. 
Integration of the equation of energy in the cross-sectional area is performed to enable a suitable introduction of the heat transfer problem solution in the one-dimensional acoustic analysis, use made of the \textcolor{black}{averaged value of temperature at each section,
\begin{equation}
    T_b^0(x)  = \dfrac{\int_0^R 2\pi y \rho u T dy}{\rho_b^0 u_b^0 \pi R^2}
\end{equation}
 with $y$ the radial coordinate of the tube and $\rho_b^0 u_b^0 = \int_0^R 2 \pi y\rho u dy/\pi R^2$ the analogous averaged momentum.} Finally, heat flux per unit area at the wall is assumed to be proportional to the temperature jump between the outer bath temperature and the section-averaged temperature of the gases $T_{\rm ext} - T_b^0(x)$. Therefore, radial conduction terms can be approximated as \textcolor{black}{$\sim \mathrm{Nu} k_b(T_u-T_b^0)/R^2$, where the temperature at the wall equals the unburnt gas temperature $T_u$, $k_b$ is the thermal conductivity of the burnt gases and $\rm Nu$ is the Nusselt number. The equation that describes the evolution of $T_b^0$ under the presented assumptions is
\begin{equation}
	\rho_b^0 u_b^0 c_{pb} \frac{\mathrm{d} T_b^0}{\mathrm{d} x} = \mathrm{Nu} k_b \frac{T_u - T_b^0}{R^2},
\end{equation}
with $c_{pb}$ the specific heat evaluated at the burnt-side temperature $T_b$. In particular, averaged values of transport coefficients $\overline{k_{b}}$ and $\overline{c_{pb}}$ are selected to represent the simplified problem over a range of burnt gas temperatures $500<T_b<1900$~K. The non-dimensional form of this equation,
\begin{equation}
	\frac{\mathrm{d} \theta}{\mathrm{d} \xi} = - \frac{\mathrm{Nu} k_b}{\rho_b^0 u_b^0 c_{pb} R} \frac{L}{R} \theta = -\sigma \theta,
	\label{eq:theta_xi}
\end{equation}
with $\theta = (T_u-T_b^0)/(T_u-T_b)$, introduces the cooling parameter $\sigma = L/l_c$, defined as the ratio between the length of the tube and the cooling length
\begin{equation}
    l_c \simeq \frac{\rho_b^0 u_b^0 c_{pb} R }{\mathrm{Nu} k_b} R = \frac{\mathrm{Pe}}{\mathrm{Nu}}R.
    \label{eq:lc}
\end{equation} 
Therefore, the parameter $\sigma$ is a function of the P\'eclet number $\mathrm{Pe}$, the Nusselt number $\rm Nu$, and the aspect ratio of the tube. In order to evaluate the cooling length, the value for the Nusselt number following \cite[Chapter~9]{kays1993convective} for laminar flows with constant wall temperature is selected to be $\rm Nu=3.66$, accordingly with the present thermally-controlled slender tube used in this work. The P\'eclet number is in turn evaluated using the values of the different physical quantities provided in Table~\ref{tab:1}. All things considered, the numerical evaluation of eq.~\eqref{eq:lc} gives a cooling distance $l_c \simeq 0.05$~m which will be used in the rest of the document to characterize the exponential temperature decay for our set of experiments in tubes of $R=5$~mm. Consequently, the cooling parameter $\sigma = L/l_c$ only changes when the length of the tube $L$ is modified.}

\begin{table}
\center
\begin{tabular}{ccccccccc}
\begin{tabular}[c]{@{}c@{}}$S_u$\\ $\mathrm{[m/s]}$\end{tabular} & \begin{tabular}[c]{@{}c@{}}$\rho_u$\\ $\mathrm{[kg/m^3]}$\end{tabular} & $\epsilon = \frac{T_b}{T_u}$ & \begin{tabular}[c]{@{}c@{}}$u_b$\\ $\mathrm{[m/s]}$\end{tabular} & \begin{tabular}[c]{@{}c@{}}$\rho_b$\\ $\mathrm{[kg/m^3]}$\end{tabular} & Nu   & \begin{tabular}[c]{@{}c@{}}$R$\\ $[\mathrm{m}]$\end{tabular} & \begin{tabular}[c]{@{}c@{}}$\overline{c_{pb}}$\\ $[\mathrm{J/kgK}]$\end{tabular} & \begin{tabular}[c]{@{}c@{}}$\overline{k_{b}}$\\ $[\mathrm{W/m\>K}]$\end{tabular}  \\ \hline
0.387                 & 1.15           & 7.37         & 2.461          & 0.156                & 3.66   & $5\times10^{-3}$    & 1120             & 0.063        
\end{tabular}
\caption{Values used to evaluate $l_c$ for methane-air flames throughout the present document.}
\label{tab:1}
\end{table}


\subsection{Non-adiabatic acoustics}

In sufficiently long tubes, $\sigma \gg 1$, experimental observations are in disagreement with predictions given by eq.~\eqref{eq:Clavin_freq} for the evolution of acoustic frequencies along the tube. Longer tubes and greater surface-to-volume ratios, for decreasing diameters, increase the impact of heat losses that produce decreasing temperature profiles in the burnt gas $T_b^0(\xi)$. This variation of burnt gas temperature with the longitudinal coordinate causes the acoustic waves propagation velocity $c_b$ to change with $\xi$ in the burnt gases region $r<\xi<1$, an important point that has been recurrently disregarded in the interpretation of experimental measurements.

Straight-forward integration of eq.~\eqref{eq:theta_xi} from a moving reference frame fixed on a steady planar flame front propagating at velocity $S_u$, provides the temperature profiles $\theta({\xi}) = \exp(-\sigma(\xi -r))$. Hereafter, the dimensionless base temperature distribution in the burnt gases can be rewritten as,
\begin{equation}
    \Delta(\xi) = \frac{T_b^0(\xi)}{T_u} = 1  + (\epsilon -1)\exp[-\sigma(\xi - r)].
\end{equation}
\begin{figure}
    \centering
	\includegraphics[width=0.495\textwidth]{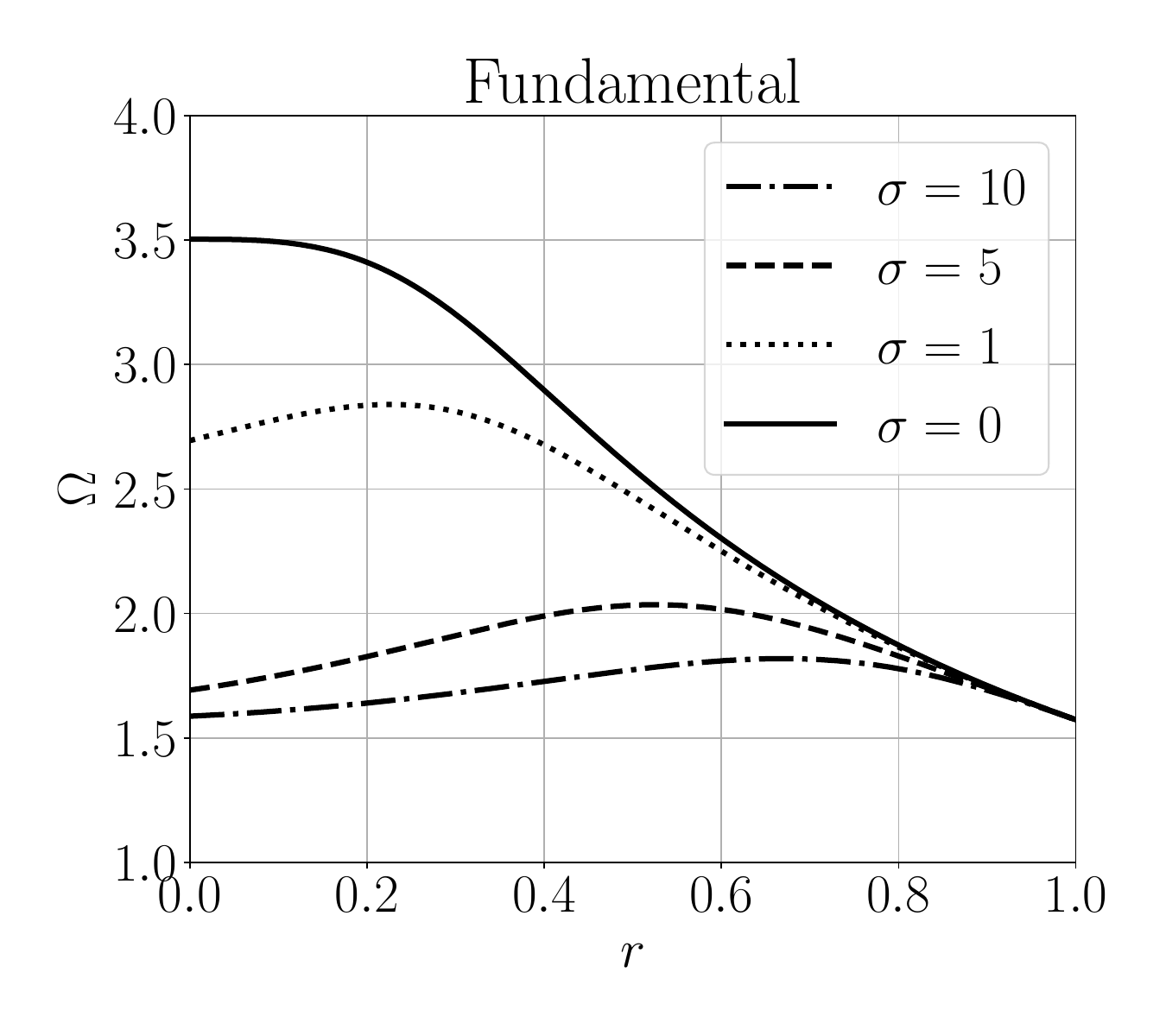}
	\includegraphics[width=0.495\textwidth]{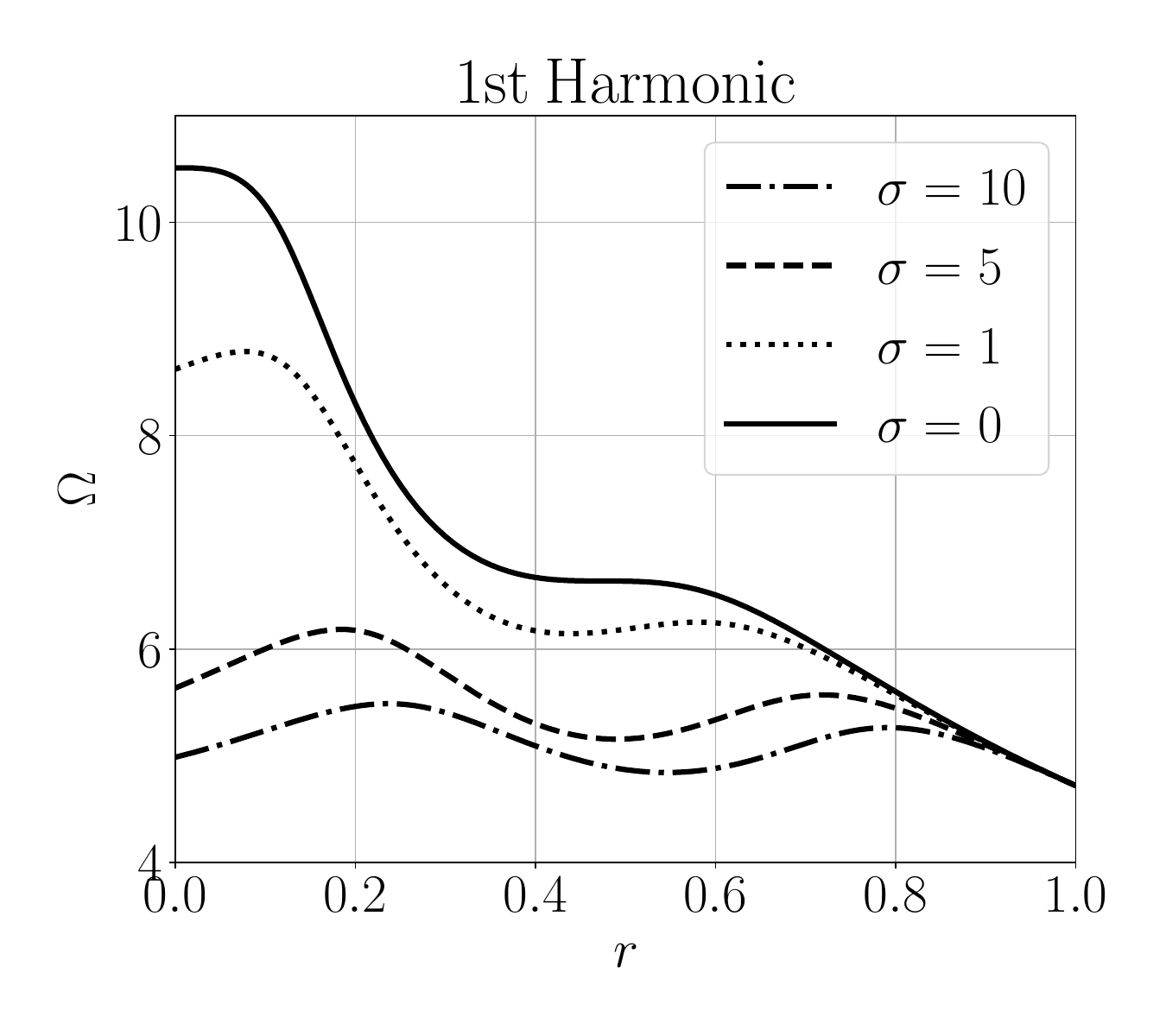}
	\caption{Non-adiabatic predictions of the dimensionless coupling frequency distributions with flame position for fundamental and fist harmonic acoustic modes. The curves for $\sigma=0$ correspond to eq.~\eqref{eq:Clavin_freq}. }
	\label{fig:fundamental_epsilon5_noniso}
\end{figure}
Introduction of burnt gas temperature distribution into eq.~\eqref{eq:wave_u} defines a non-isothermal acoustics problem. In particular, some analytic solutions exist for exponential temperature variation laws $\Delta(\xi)$ \citep{musielak2006method}. Nevertheless, for each $0<r<1$ flame position, the eigenvalues of the system of partial differential equations, with open and closed end boundary conditions \eqref{eq:bcs}, and jump conditions at the flame \eqref{eq:jcp}-\eqref{eq:jcu} can be numerically computed to offer a more accurate description to the oscillation frequency than eq.~\eqref{eq:Clavin_freq}. The resulting evolution with $r$ of the acoustic coupling frequencies of the tube $\Omega$ depends on both the temperature ratio $\epsilon$ and the cooling parameter $\sigma$. Figure~\ref{fig:fundamental_epsilon5_noniso} shows the evolution of the dimensionless acoustic frequency $\Omega$ with $r$ for $\epsilon=5$ and different values of $\sigma$ in the fundamental mode and first harmonic. The original adiabatic solution $\sigma=0$, depicted with solid lines, can be compared to long temperature decay distances $l_c \gg L$. Oppositely, increasing heat-loss effect flattens the distribution by providing a burnt-side temperature that is rapidly cooled down to the wall temperature $T_{\rm ext} = T_u$ and, thus, an isothermal gas condition is approached for $\sigma \gg 1$. 

In order to compare the new non-isothermal acoustics model with experimental data a flame tube is placed in a thermal bath that maintains the wall temperature to a fixed value by means of a recirculating water flow. Temperature control is key to the non-adiabatic validation and will be shown to play a major role in controling the transitions between primary and secondary thermoacoustic instabilities.

\section{Experimental setup and procedure}
\label{sec:experimental_procedure}

\begin{figure}
    \centering
	\includegraphics[width=0.8\textwidth]{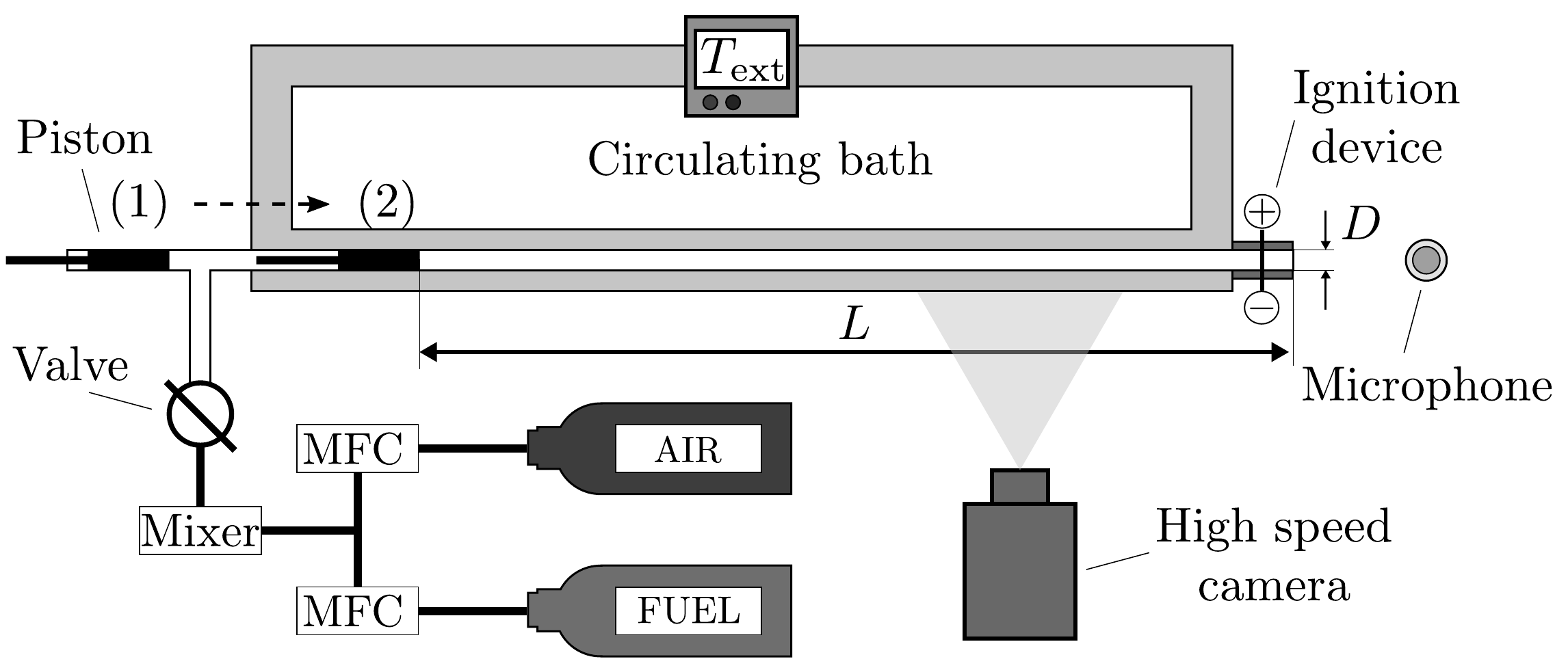}
	\caption{Schematic representation of the experimental apparatus. During the feeding phase, piston is in position $(1)$. Upon complete charge, the piston is moved to position $(2)$ enclosing an effective distance $L$.}
	\label{fig:EXSET_overview}
\end{figure}

 A simplified scheme of the experimental setup, which consists of two concentric horizontal tubes of lengths {$150$~cm}, is presented in {Fig.~\ref{fig:EXSET_overview}}. Combustion takes place inside the inner tube, which is made of borosilicate glass, has an inner diameter {$D=10$~mm}, and a wall thickness of {2~mm}. The outer tube is made of acrylic glass, has an outer diameter of {30~mm}, and a wall thickness of {3~mm}. In the space between the two tubes a constant flow of a mixture of ethylene glycol and water is established. Its temperature, $T_{\rm ext}$, is controlled by means of a Fisher Scientific Isotemp 4100~R20 refrigerated circulating bath. This temperature is varied between $273$ and {$343$~K}, where the upper value is limited by the structural integrity of the outer tube.

 In the following experiments, a flame propagates from one end of the tube, which is maintained open to the atmosphere, to the other, that is kept closed. The flammable mixture is prepared using air and $99.5$-pure methane. The equivalence ratio is controlled by adjusting the mass flow rates of air and reactant with two EL-Flow Bronkhorst mass flow controllers. The ignition device is situated at the open end and consists of a piezoelectric spark generator which is connected to two electrodes facing each other diametrically inside the tube. When the spark gap is driven, the electric arc between the two electrodes supplies the required energy to ignite the mixture. Finally, the opposite end of the tube is closed with a moveable piston that allows testing different effective tube lengths $L$. Taking all this into account, the described setup allows to study the behaviour of thermo-acoustic instabilities of the flame under variations of three parameters: equivalence ratio $\phi$, wall temperature $T_{\rm ext}$, and length of the tube $L$.

The propagation of the flame inside the tube is recorded from the orthogonal direction to the tube axis by means of a Phantom VEO~710 high-speed camera with a resolution $1280\times800~\mathrm{pixels}$. \textcolor{black}{A frame rate of 2000 fps was chosen for most of the experiments to provide a good compromise between sufficient time resolution and image brightness.} The images are used to obtain the position and velocity of the flame. Each frame of the recorded video registers the instantaneous luminosity of the flame which gives its position inside the frame. The displacement between subsequent frames and the camera frame rate enable the computation of the instantaneous velocity of the front. Due to the slenderness of the combustion chamber and the need of spatial resolution across the flame front, only partial longitudinal visualization of the setup is possible during each experiment. A field of view of approximately $200\>\mathrm{mm}$ has been selected to provide good compromise between resolution and observational length. 

Images of the flame front propagation are post-processed using an in-house Python code. Therefore, flames are discretized in the transverse direction by the pixel-row resolution of the image. Typically, a number of $64$ rows fall within the horizontal band described by the flame in its movement. The axial displacement of every point of the flame front is computed by using the cross-correlation of \textcolor{black}{every pixel row between consecutive frames}. In particular, the cross-correlation is performed over the derived image intensity values with respect to $x$ in order to prevent the flame tail from introducing a bias in the correlation and therefore a drift in flame position tracking, see \textcolor{black}{Appendix}~\ref{appendix:postprocess}. The position of the maximum value of the cross-correlation is the most probable displacement of the flame in the selected row. The accuracy of the method is improved by using a parabolic subpixel interpolation. This process is repeated for every row that contains the flame and for every pair of frames. The calculated displacement of the flame front is averaged in the vertical direction and is divided by the time lapse between two frames, which is kept constant for every video. Therefore, the velocity of the averaged flame front is obtained as a function of time. Finally, \textcolor{black}{the frequency components} of this signal are obtained by applying Fourier analysis.

Additional instrumentation includes a microphone to capture the acoustic emission of the flame during its propagation. The microphone is placed at a fixed position at the outlet of the combustion chamber and samples pressure oscillations at a $44.1\>\mathrm{kHz}$ frequency with a $16$ bit-depth. Although audio signals do not allow us to determine the absolute magnitude of pressure, they offer qualitative information about the amplitude of pressure variations and are suitable for time-frequency analysis.

Finally, the operation procedure is always performed as follows. Prior to each experiment the piston at the closed end is moved backwards allowing the mixture to enter the tube. During the feeding phase, combustion products from previous runs are replaced with fresh gases. The venting occurs at the open end of the tube. Once the tube is filled with fresh mixture, the gas feeding system is cut off by closing the valve at the inlet of the tube. In order to prevent the diffusion of the reactive mixture, which would cause problems of ignition, the open end is blocked. The effective chamber length is then adjusted by placing the piston at a distance $L$ from the open end.
Upon completion of the filling process, the mixture is allowed to settle for about one minute to ensure that the temperature of the reactants mixture equals to the refrigerated bath circulator's temperature, $T_{\rm ext} = T_u$. Then, the blockage of the ignition end is removed and the mixture is ignited \textcolor{black}{manually, with a delay of the order of 3~seconds}. 

It should be emphasized that the experimental setup enables the control of thermal boundary conditions and the unburnt gas temperature $T_u$. Therefore, the evolution of temperature downstream of the flame can be predicted from the heat-transfer analysis presented above. This is of the uttermost importance in the experimental validation of thermoacoustic frequency analysis with heat losses presented next.  

\section{Experimental results}
\label{sec:exp_res}

In this section, the experimental data of premixed flames undergoing thermoacoustic instabilities in horizontal tubes of different lengths $L$, and thermal bath temperatures $T_{\rm ext} = T_u$, is presented and discussed. Under these conditions, the computed thickness of methane flames is $\delta_T= k /(\rho c_p S_u) \simeq 5\times 10^{-2}\>\mathrm{mm}$. Therefore, the different tested tube lengths expressed in terms of flame thickness are approximately $L/\delta_T \sim \mathcal{O}(10^4)$, considering the ranges of $L=30-150$~cm, so that the acoustic wavelengths are indeed much larger than the reacting sheet thickness.

\subsection{Phenomenology}
\begin{figure}
    \centering
	\includegraphics[width=1\textwidth]{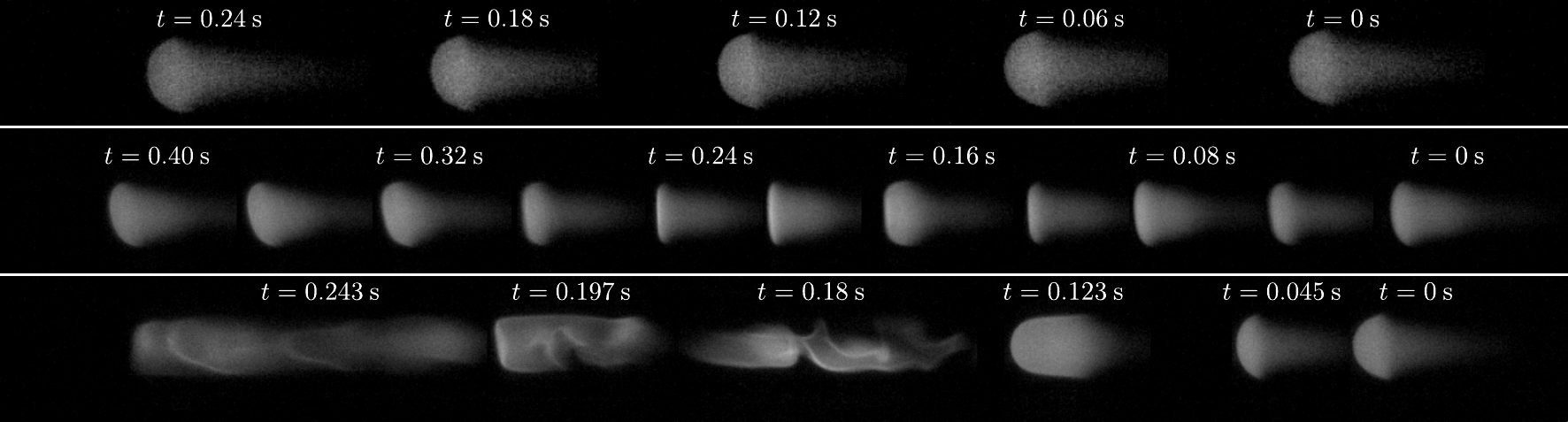}
	\caption{Flame-front propagation snapshots along the experimental tube for three regimes: Steady propagation (top row); primary instabilities (mid row); and secondary instabilities (bottom row). }
	\label{fig:flame_evolution}
\end{figure}

Three main propagation regimes have been identified in the experiments through high-speed video recording. Figure~\ref{fig:flame_evolution} summarizes the features of each regime in terms of front shape and displacement. The first regime is characterized by the absence of thermo-acoustic instabilities, where a steady propagation of the flame front is observed for equispaced snapshots every $0.06$ seconds on the top row, propagating from right (open end) to left (closed end). The steady propagation velocity is higher than the planar flame velocity $S_L$, as expected from curved fronts far from the quenching limit. Moreover, the pulsating flame regime shown in the mid row of Fig.~\ref{fig:flame_evolution} is associated to primary instabilities. In this regime, flames are known to oscillate with a rather constant amplitude and with significant front curvature reduction, which can be noted as the flattening of the flame. Finally, the snapshot composition in the bottom row of Fig.~\ref{fig:flame_evolution} depicts the flame motion under secondary instabilities. Under certain conditions, the flame front transitions from primary to secondary instabilities. First, the previous nearly-planar front is perturbed with a smooth wrinkling of the surface. Then, the amplitude of this corrugation grows fast and soon the flame becomes a folded and stretched disordered surface. During this phase, the front \textcolor{black}{loses} its coherence and propagates faster due to the large increase in burning area. The absence of a well-defined flame surface complicates the tracking of its position and the measurement of the propagation velocity during this phase. Thus, once the secondary instability is fully developed, only a qualitative description of flame front propagation is possible from recorded images. In particular, various transitions between regimes can take place in the same tube during the complete propagation of the flame.

It should be pointed out that the flame velocity in the laboratory fixed reference frame, $u_f$, differs a small amount from the flame propagation velocity $S_u$ because of the nearly-stagnant flow condition found in the mixture of reactants before the flame front. Nevertheless, acoustic oscillations trigger a small amplitude motion in the fluid that needs to be accounted for if specific values of flame propagation velocities are required. 
 
\begin{figure}
 \centering
    \includegraphics[width=1\textwidth]{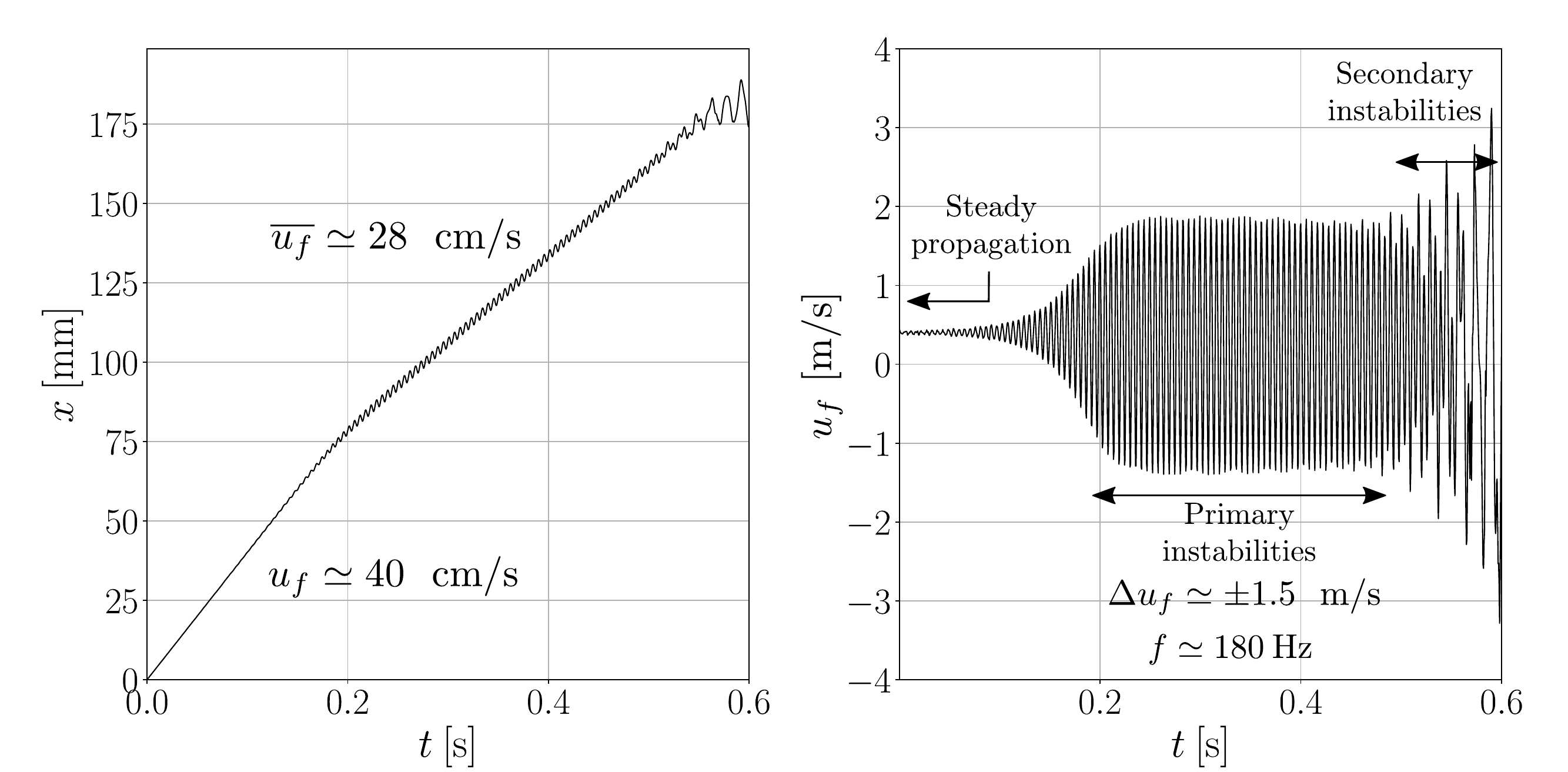}
 \caption{Position and velocity of the flame front against time obtained from video recording of a methane-air mixture with  $\phi=1.3$, $L=150$~cm and $T_u = 323\>\mathrm{K}$. }
 \label{fig:velocityandposition}
\end{figure}

Experimental results of flame position and velocity with time in a {$L=150$~cm} tube, for a rich methane-air mixture of equivalence ratio $\phi = 1.3$ and wall temperature $T_u = 323$~K are presented in Fig.~\ref{fig:velocityandposition}. An initial constant-speed motion is captured after ignition, followed by the initiation of smooth oscillations that are representative of primary thermoacoustic instabilities, which develop through an exponential growth of pressure and velocity oscillations, following unstable linear-system behavior. At some point, nonlinear effects and damping mechanisms become important, causing the exponential growth to decay and oscillation amplitude saturation around a fixed value. Differences in terms of flame velocity between steady propagation and averaged primary oscillations are noticeable. \textcolor{black}{Specifically, the averaged flame velocity during the pulsating flame regime, where the flame surface and curvature are reduced, is very similar to the theoretical planar flame velocity $S_L = 0.273$~m/s in this experiment, and lower than the steady propagation speed under the initial curved front. Flame flattening precedes the appearance of small corrugations at the front, which then grow fast increasing the reacting surface area}. This gives rise to the large-amplitude secondary instabilities that rapidly produce complex reacting-front surfaces, which become difficult to track through the high-speed video images \textcolor{black}{and develop for $t>0.55$ s in Fig.~\ref{fig:velocityandposition}}.

\begin{figure}
    \centering
	\includegraphics[width=0.9\textwidth]{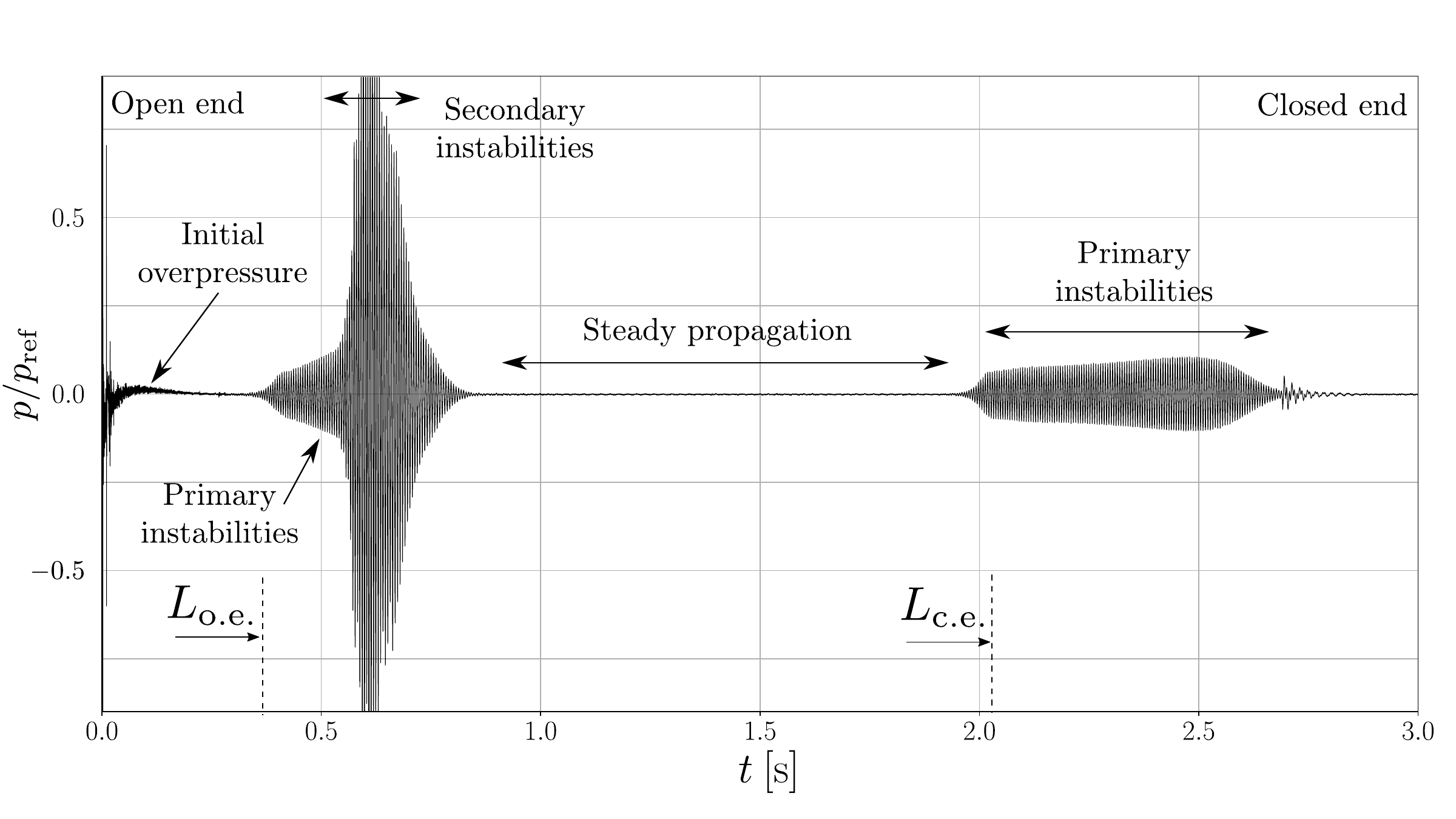}
	\caption{Audio signal recorded for $L =150$~cm, $T_u=313$~K and $\phi =0.95$. The propagating flame undergoes primary oscillations and transitions to the secondary regime at distance $L_{\rm o.e}$ before restabilizing at the mid part of the tube, to finally excite primary instabilities near the closed end at distance $L_{\rm c.e.}$.}
	\label{fig:Audio_signal}
\end{figure}

Nonetheless, the audio signal provides extra quantitative information when the video processing fails upon arising of secondary oscillations. In particular, the acoustic oscillation frequency can be measured during the secondary instability regime and along the whole length of the tube for each run, which becomes a difficult task for imaging methods in such slender configurations. Figure~\ref{fig:Audio_signal} shows the complete audio signal in time for $L = 150$~cm, $\phi = 0.95$ and $T_u =313$~K. This measurement is proportional to acoustic pressure, which allows to link propagation regimes to the order of magnitude of pressure oscillations in the tube with a common reference pressure $p_{\rm ref}$ fixed through the microphone setup. In this particular experiment, the flame experiences the evolution from steady propagation to primary instability close to the open-end ignition event. Soon, transition from primary to secondary instability occurs and the flame front becomes strongly corrugated, where the amplitude of pressure oscillations shoots up. Later, re-stabilization occurs in the middle region of the tube, flame front coherence is recovered, and propagation remains stable and steady. Finally, as the flame approaches the closed end, a second appearance of primary instabilities occurs. This time, no secondary instability develops and, after a series of cycles, flame front re-stabilizes.

This rich variety of phenomena is affected by modifications in tube length, bath temperature, and equivalence ratio. In particular, it will be \textcolor{black}{shown} that typical transitions in far-from-stoichiometric mixtures produce flame extinction, while reactions of nearly-stoichiometric mixtures can withstand the strain produced by high pressure oscillations and recover the steady propagation regime after developing disordered patterns. Furthermore, flames do \textcolor{black}{not} transit from primary to secondary instabilities under certain combinations of the controlling parameters, \textcolor{black}{for} particularly \textcolor{black}{long} tubes and increasing temperatures. 

\subsection{Acoustic mode variation with tube length}

A variation of the length of the combustion volume is offered through progressive displacement of the piston at the closed end of the tube. This allows modifying the acoustic time involved in the wave displacement along a distance $L$. Furthermore, the extended frequency analysis for non-adiabatic tubes is controlled by the relation of tube length to cooling distance $\sigma = L/l_c$, which renders nearly-isothermal solutions in long tubes $L\gg l_c$, and nearly-adiabatic solutions in the short-tube limit $L\ll l_c$.

\begin{figure}
 \centering
	\includegraphics[width=1\textwidth]{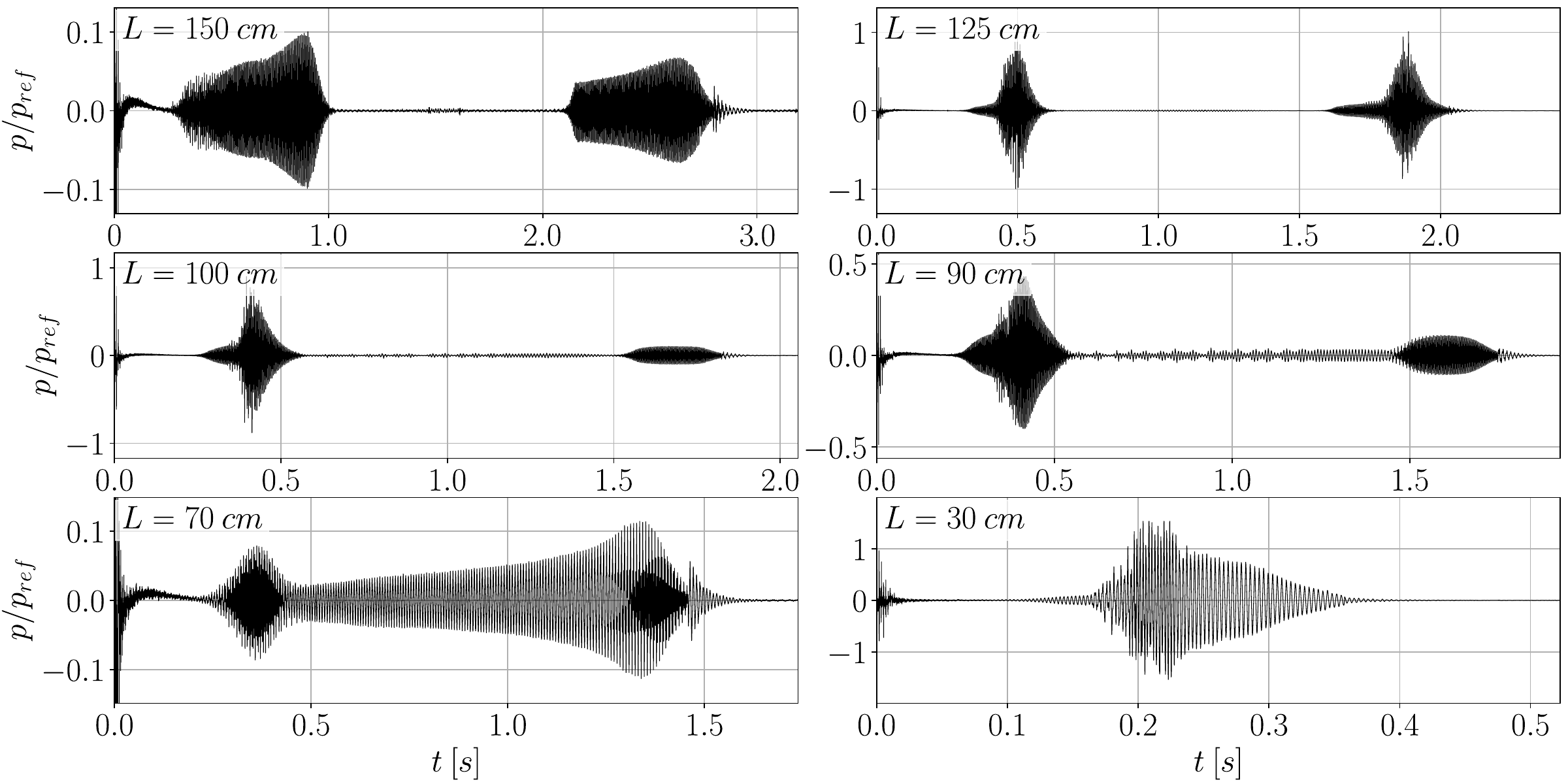}	
 \caption{Audio signals from experiments with different tube lengths $L$, with $T_u = 303$~K and $\phi = 1$, and constant amplitude reference measurement $p_{\rm ref}$. \textcolor{black}{First emerging oscillations correspond to the $L_{\rm o.e}$ region, followed by a second stage of oscillations in the $L_{\rm c.e.}$ region, except in the $L=30$ cm case. }}
 \label{fig:audio_signal_length}
\end{figure}
 
First results of the experimental tests can be extracted from Fig.~\ref{fig:audio_signal_length}, that shows the measured acoustic emission for decreasing tube lengths with stoichiometric mixtures, $\phi = 1$, and bath temperature equal to the unburnt gases temperature, $T_u = 303$~K. Acoustic pressure variations are referred to a reference amplitude value $p_{\rm ref}$, which is chosen in such a way that secondary instabilities are of order unity and primary oscillations one order of magnitude smaller. In every panel showing audio signals, microphone position and gain are equal so that the pressure scale is common and therefore amplitudes can be compared noting the span of the vertical axes.

In very long tubes, $L = 150$~cm, two excited regions undergoing primary instabilities exist around certain sections of distances $L_{\rm o.e.}$ and $L_{\rm c.e.}$ measured from the closed end, where {o.e.} and {c.e.} stand for open and closed end respectively, and indicate the nearest end of the tube to the corresponding instability. Depending on the experimental conditions, the distance of the section near the open end $L_{\rm o.e.}$ ranges from $3L/4$ to $4L/5$, and the one next to the closed end $L_{\rm c.e.}$ is approximately between $L/4$ and $L/5$. Later on, it will be shown that these oscillations correspond to the destabilization of the first harmonic mode and that other regions are of interest when considering the excitation of the fundamental mode.

Furthermore, the amplitude of pressure oscillations displays a growing trend with length reduction \textcolor{black}{from $150$~cm to $100$~cm}. Tubes of $L=125$~cm show double development of secondary thermoacoustic instabilities at both coupling regions. These lead to pressure oscillations whose amplitude is typically an order of magnitude higher than those observed for primary instabilities, easily identified in the $L=100$~cm and $L=90$~cm panels of Fig.\ref{fig:audio_signal_length}. As the tube length is reduced to $L=100$~cm, the second transition is eliminated and weak oscillations of a lower frequency appear in between these two coupling regions. This new mid-tube regime is in turn associated to the fundamental mode.

Further reduction to $L = 70$~cm displays a change in the unstable behaviour of acoustic modes. The oscillatory motion of the flame presents here two differentiated frequency components associated to the fundamental and first harmonic modes. Broadly speaking, fundamental mode instability seems to be more intense in the central region while first harmonic destabilization seems to be preferential around $L_{\rm o.e.}$ and $L_{\rm c.e.}$ regions. Finally, in shorter tubes, $L<60$~cm, a sole unstable region exists. The oscillations observed in shorter tubes correspond to the destabilization of the fundamental mode, which always leads to secondary instabilities under these temperature, diameter and equivalence ratio conditions.

\begin{figure}
    \centering
	\includegraphics[width=0.45\textwidth]{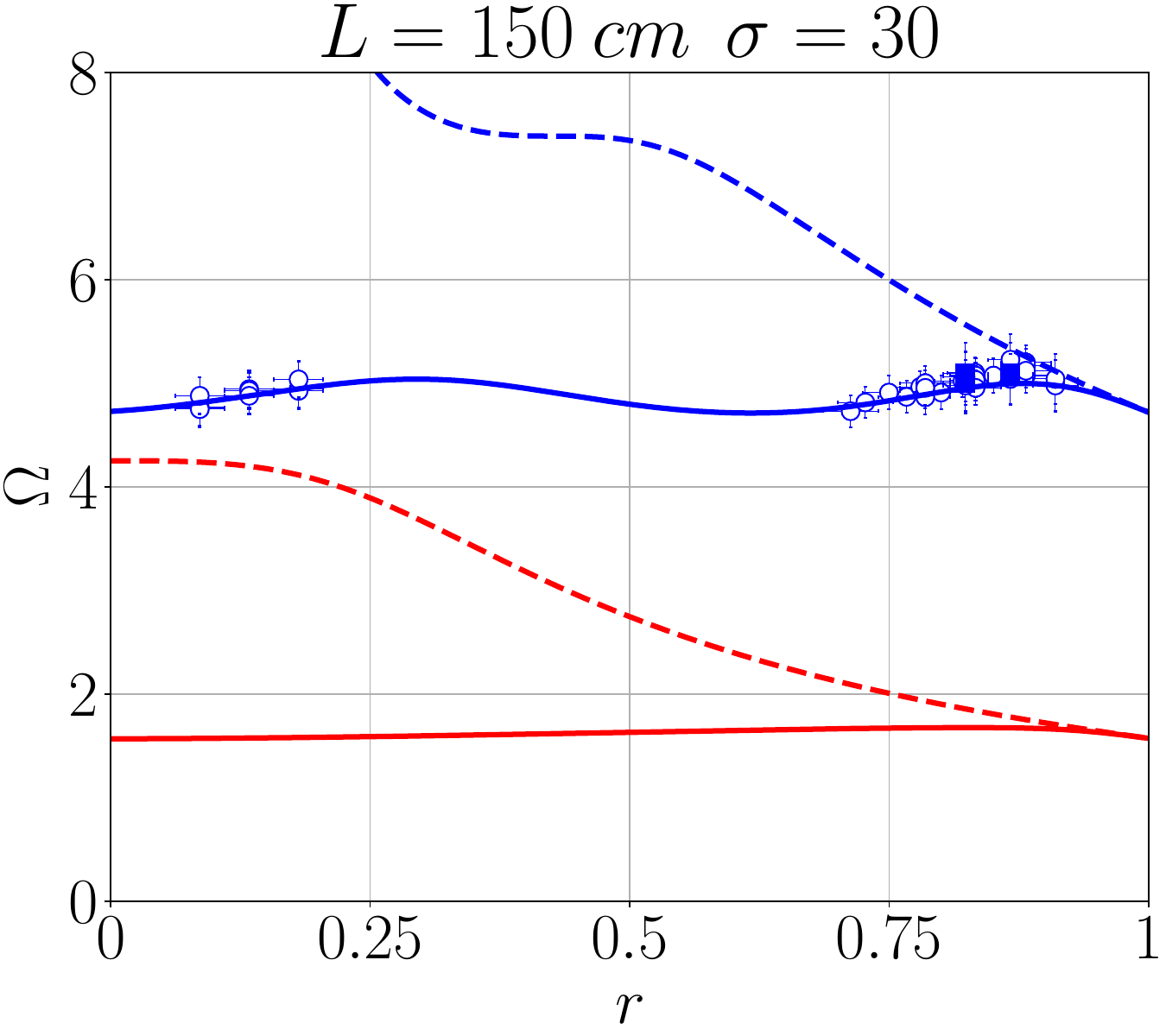}
	\includegraphics[width=0.45\textwidth]{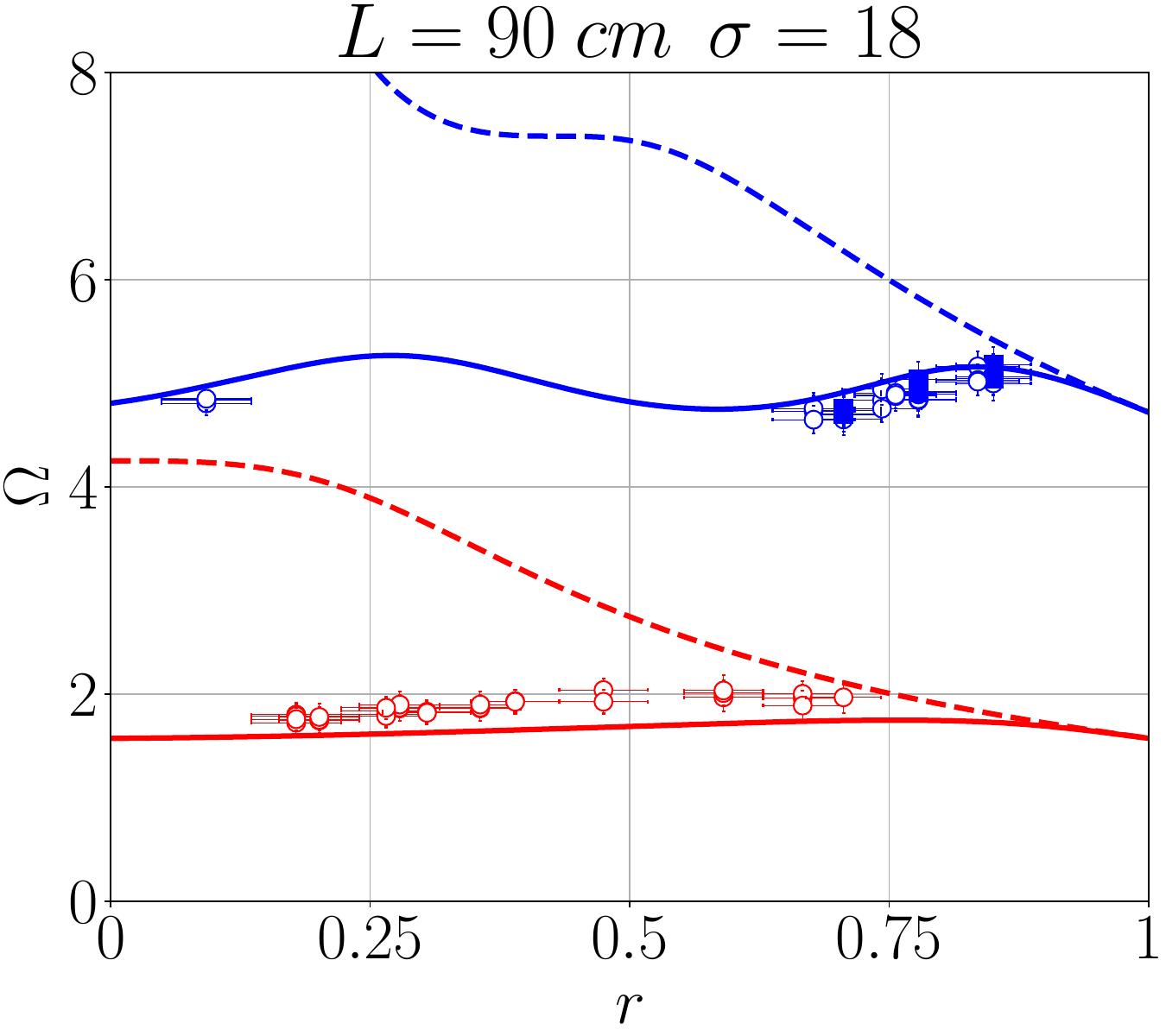}
	\includegraphics[width=0.45\textwidth]{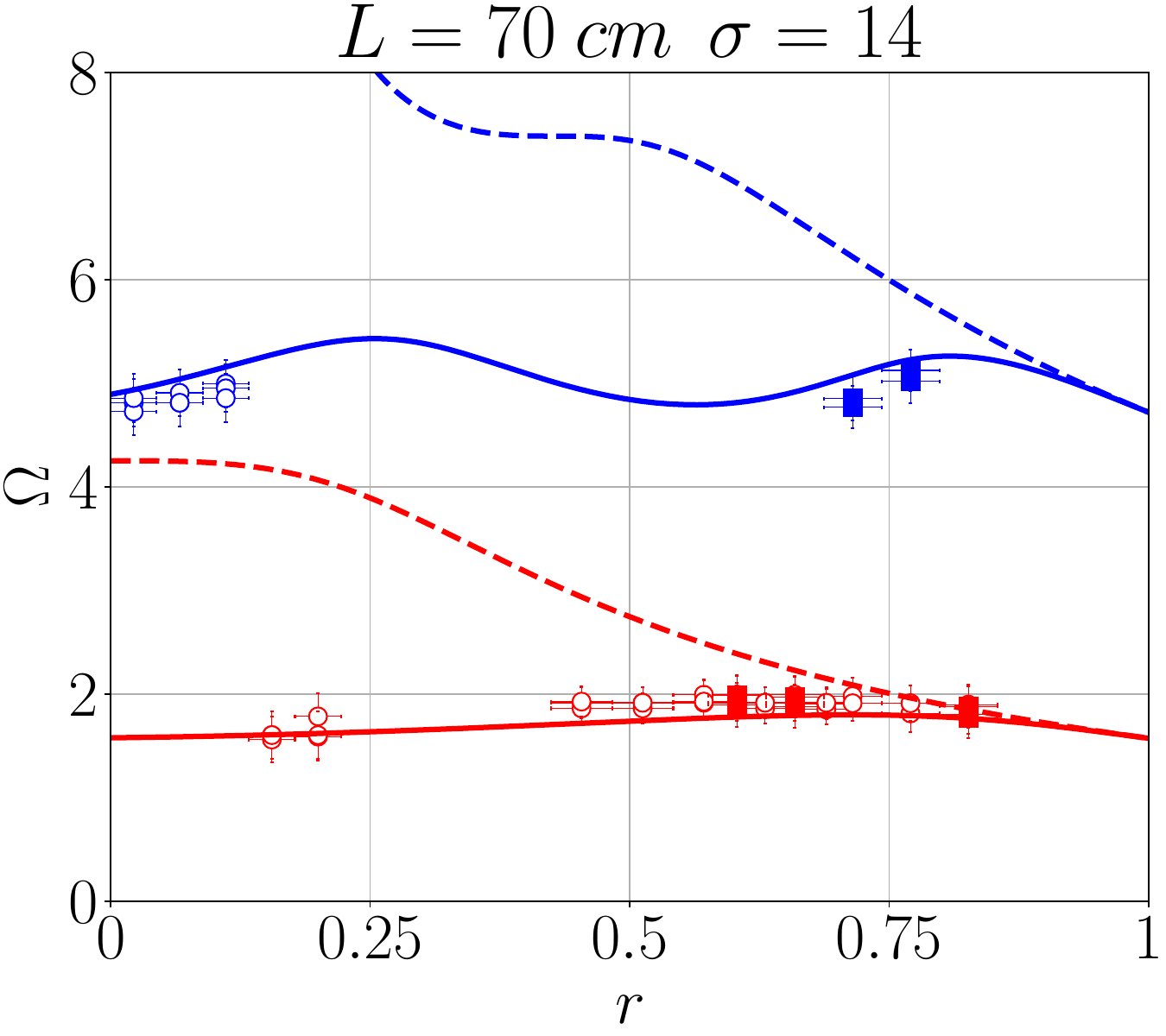}
	\includegraphics[width=0.45\textwidth]{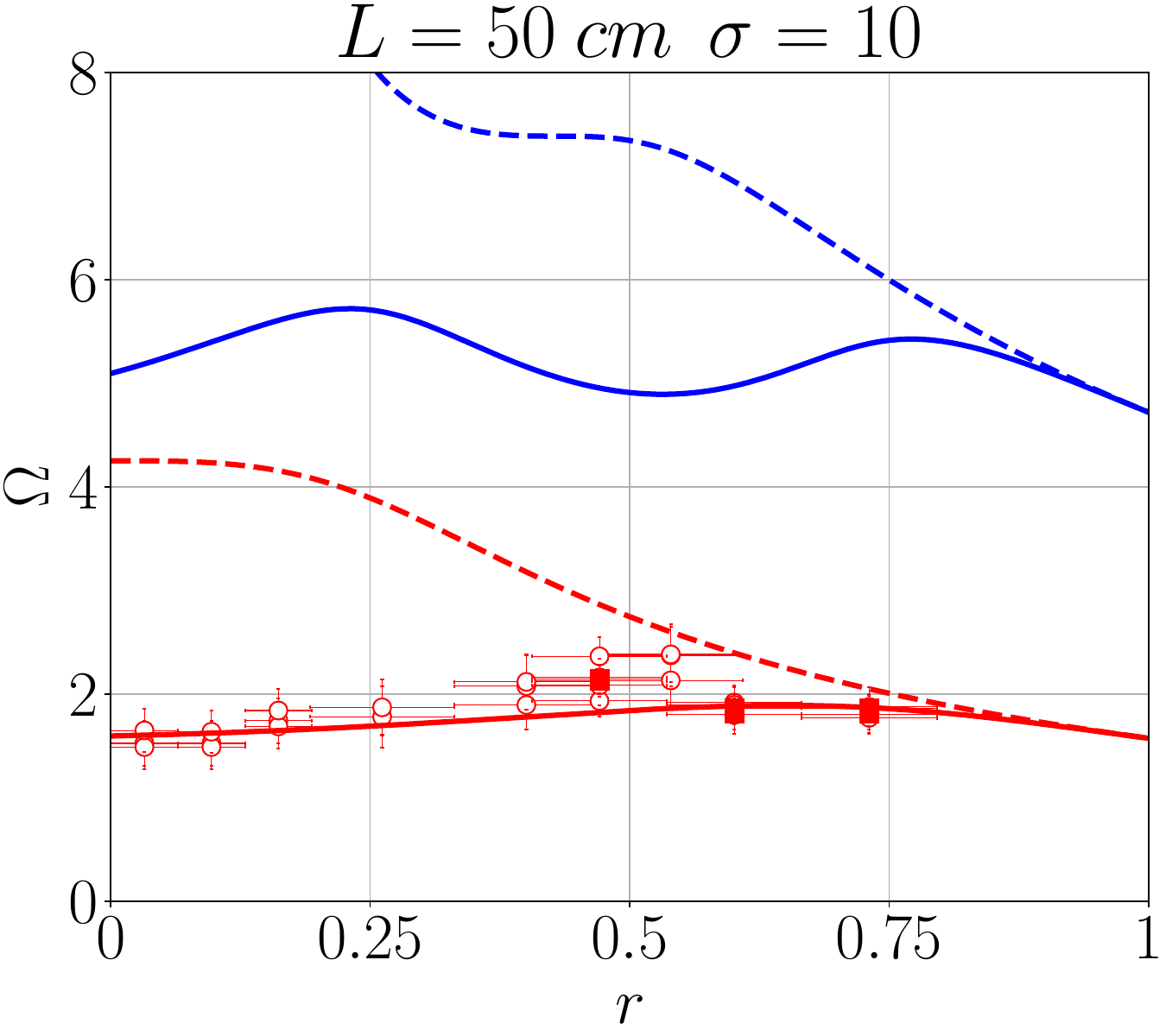}
	\caption{ Theoretical predictions of the dimensionless coupling frequency (solid lines) for the fundamental (red) and first harmonic (blue) modes along the tube {using the extended non-adiabatic model}, and experimental data (symbols) at $303$~K (empty circles) and $333$~K (solid squares). Dashed lines correspond to the one-dimensional adiabatic model of eq.~\eqref{eq:Clavin_freq}.}
	\label{fig:video_frequencies}
\end{figure}

For the sake of completeness, video processing of the flame front can provide here key information regarding the base frequency of the thermoacoustic dynamics in different regions of the tube. In particular, Fig.~\ref{fig:video_frequencies} shows the comparison between the theoretical frequencies obtained by the adiabatic prediction (dashed curves), extended non-adiabatic model (solid curves) and experimental flame image oscillation frequencies (symbols). There, fundamental mode (red) and first harmonic frequencies (blue) are identified. Moreover, the use of dimensionless frequency, $\Omega = \omega t_a$, enables the collapse of predictions for variable temperature conditions, as the reference acoustic time, \textcolor{black}{$t_a =L/c_u$}, is modified via the main sound speed of the mixture. Empty circles correspond to experiments with $T_u= 303$~K and filled squares are experimental data with $T_u = 333$~K, where the \textcolor{black}{error bars represent} the experimental uncertainty \textcolor{black}{as detailed in Appendix~\ref{appendix:postprocess}}.

It is now recalled that video recording in such slender configurations implies numerous tests to cover the imaging of the tube by composition of short longitudinal sections of around $20$~cm, depicted by the collection of symbols. However, good agreement of the frequency theoretical prediction and experiments is also obtained when considering the audio signals that provide a continuous frequency description along the whole tube, and of the same value of image oscillation frequencies.

Firstly, it is clear that \textcolor{black}{increasing tube lengths provide greater values of $\sigma=L/l_c$ with constant cooling length and, therefore, more pronounced modifications to the adiabatic predictions given by eq.~\eqref{eq:Clavin_freq}. Specifically, theoretical estimation of the cooling length $l_c$ is recalled here to yield a value of $5$~cm. This estimation has been used throughout the whole document to compare experimental results and theoretical predictions in the $10$~mm diameter} tube used here. The experimental results are in close agreement with the heat-loss model presented here, which allow to \textcolor{black}{identify which acoustic modes are involved in the oscillation}. The fundamental mode (red) is shown to dominate for shorter tube lengths, contrarily to the first harmonics (blue), that become more important with increasing lengths. Although first harmonics near the closed end of the tube ($r\simeq0.1$) could be mistaken for adiabatic fundamental modes (red dashed curve) if taken separately from the experiments, the rest of the measurements providing the frequency distribution along the tube should then be identified as inconsistent with the adiabatic model. 

\subsection{Bath temperature effects}

\begin{figure}
    \centering
    \includegraphics[width=0.475\textwidth]{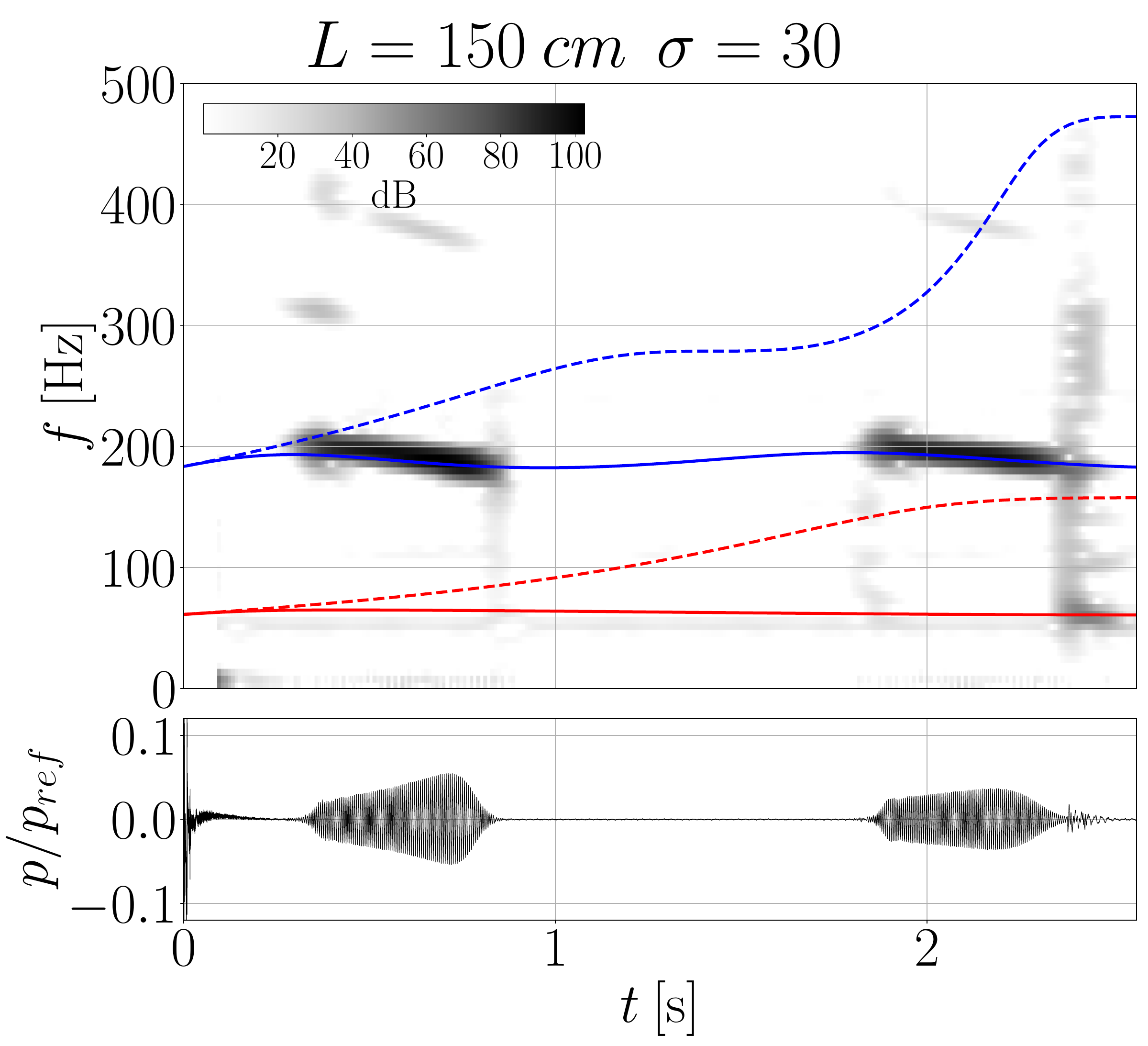}
    \includegraphics[width=0.475\textwidth]{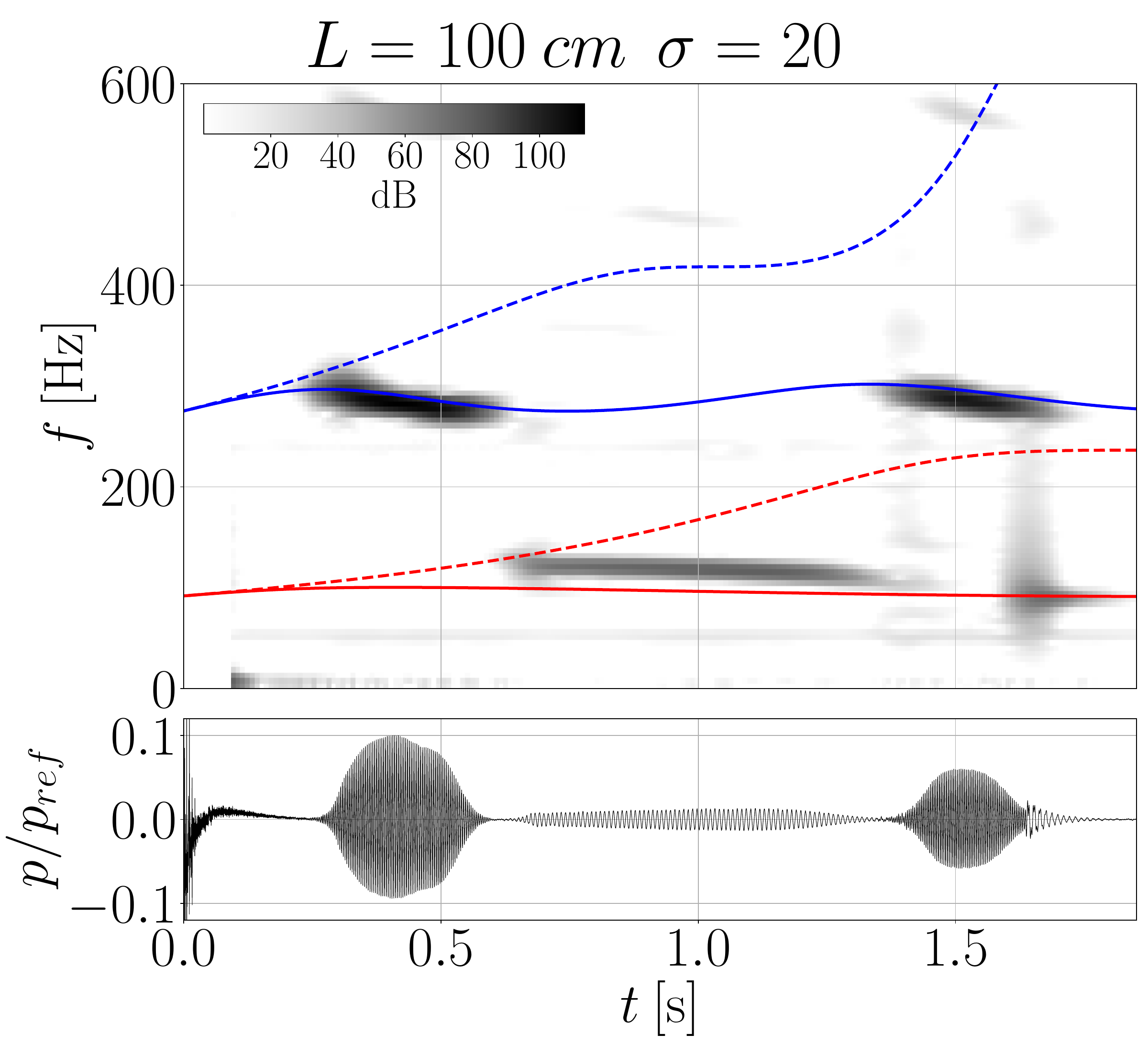}
    \includegraphics[width=0.475\textwidth]{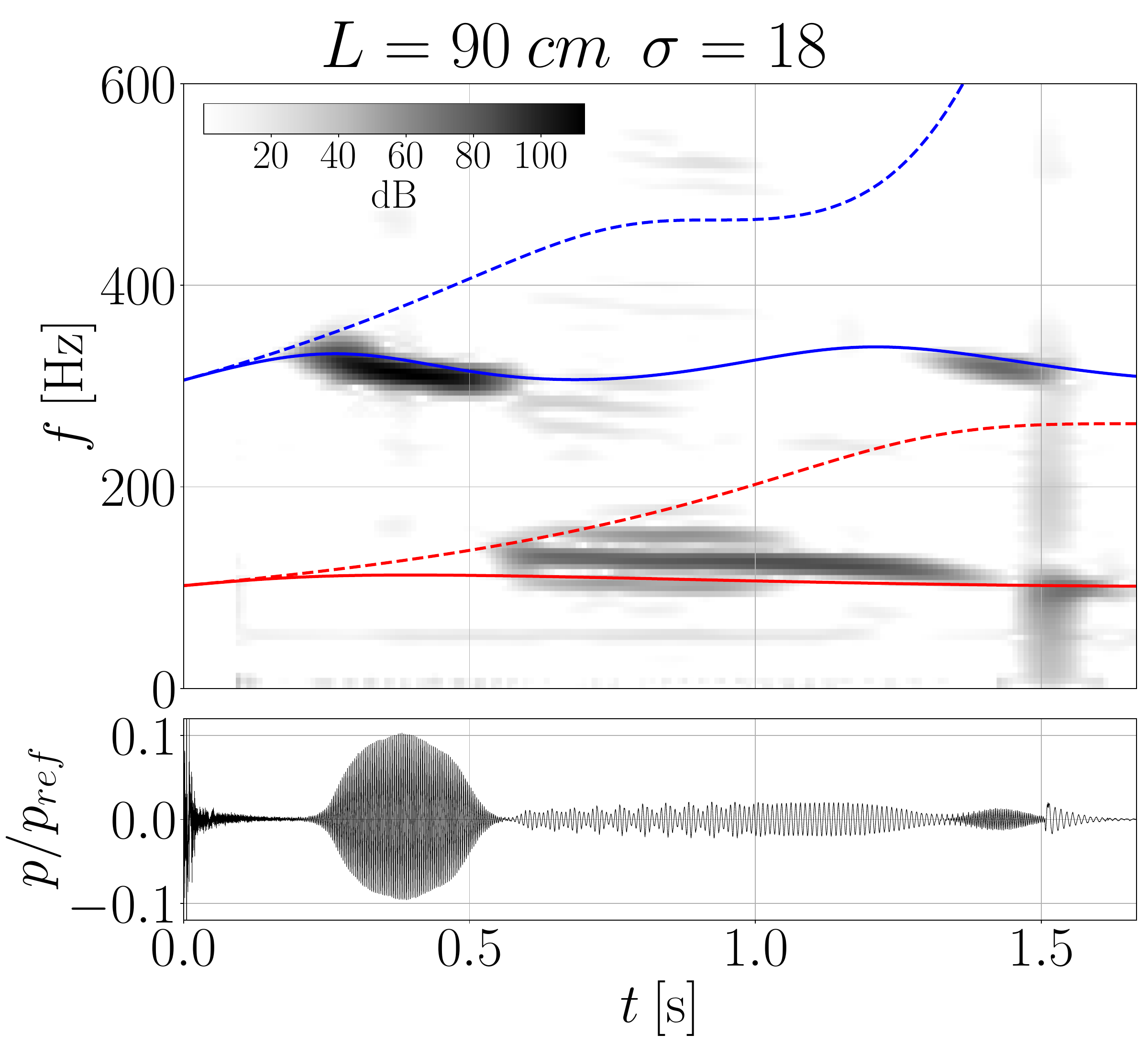}
    \includegraphics[width=0.475\textwidth]{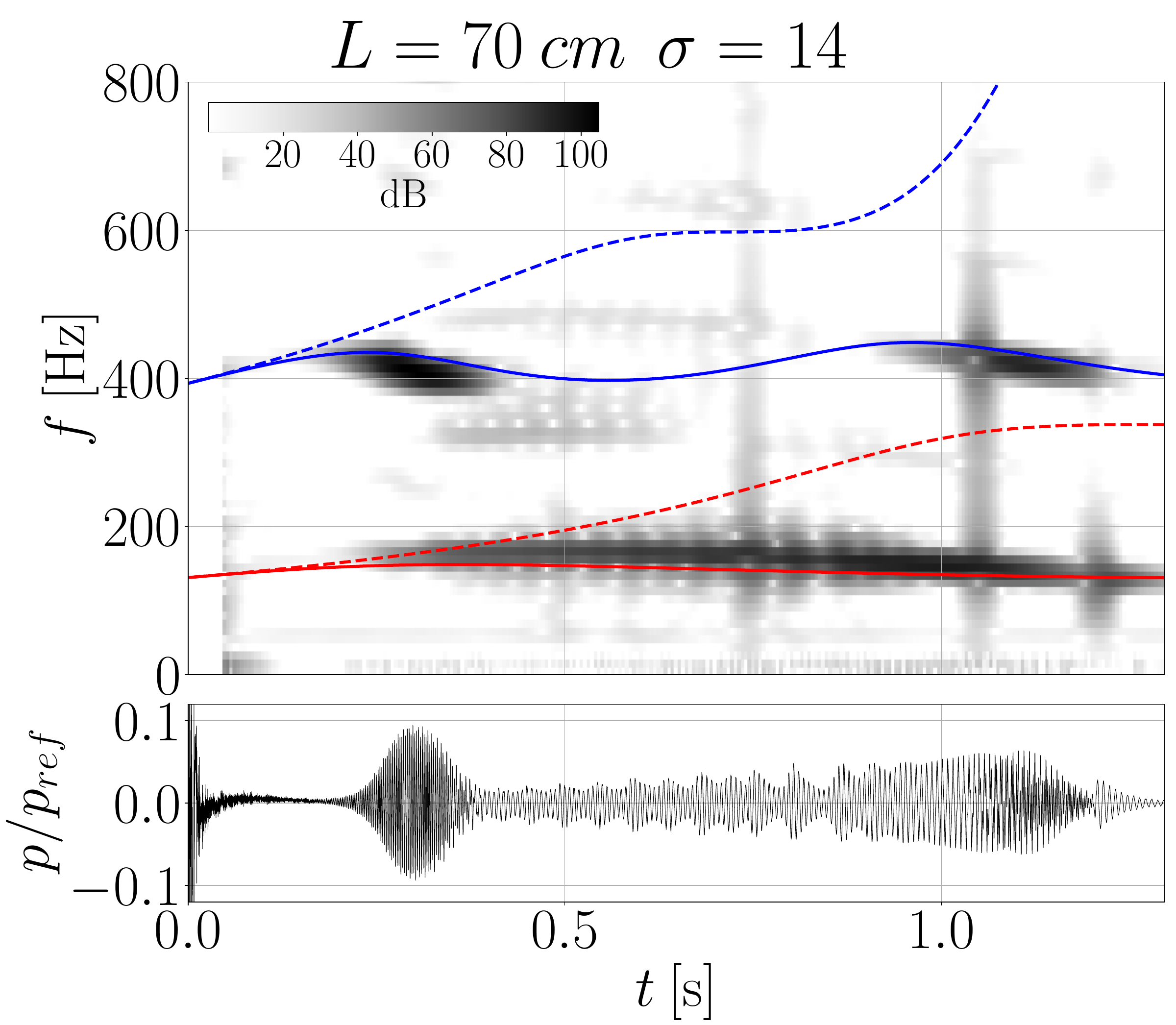}
    \includegraphics[width=0.475\textwidth]{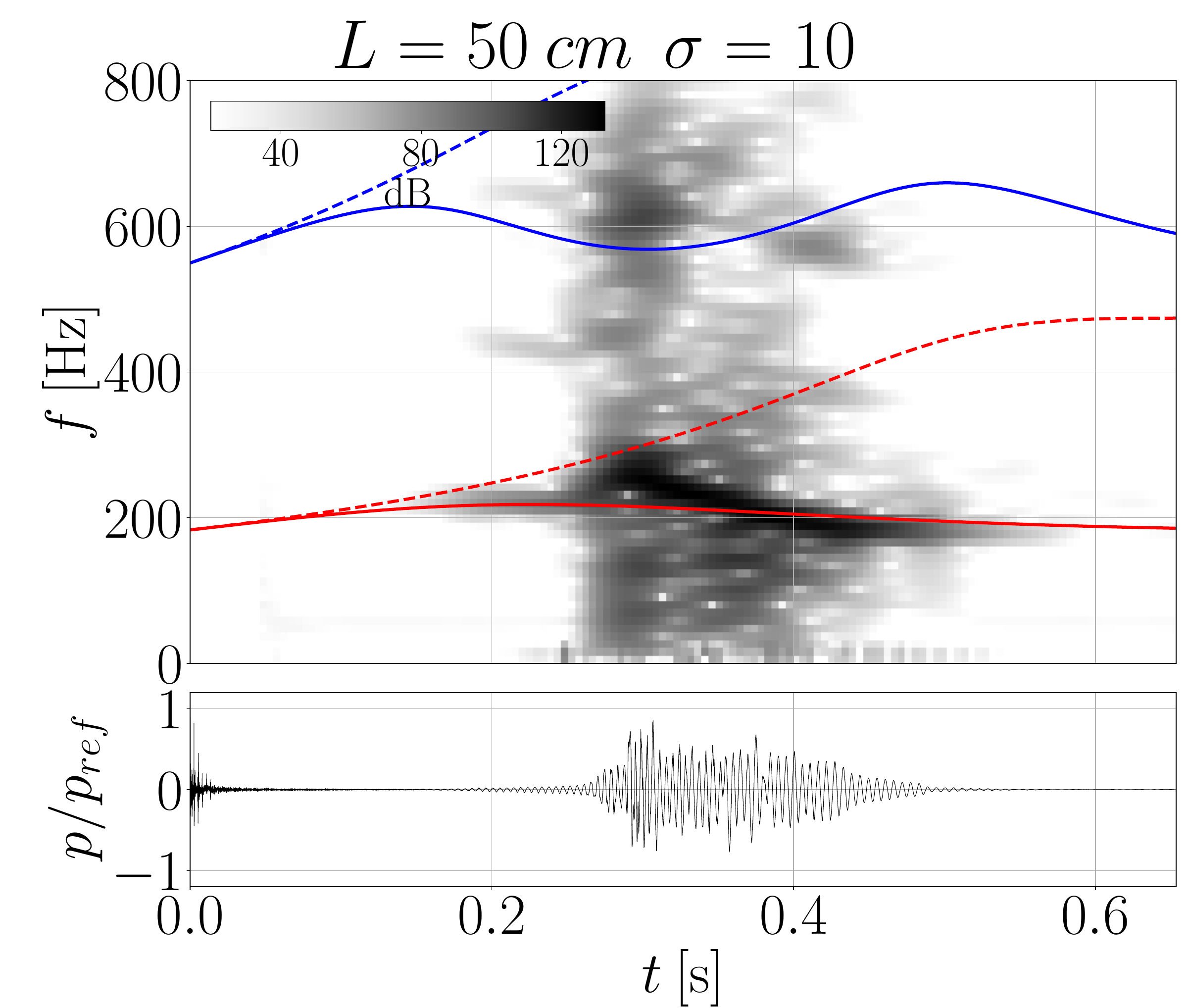}
    \includegraphics[width=0.475\textwidth]{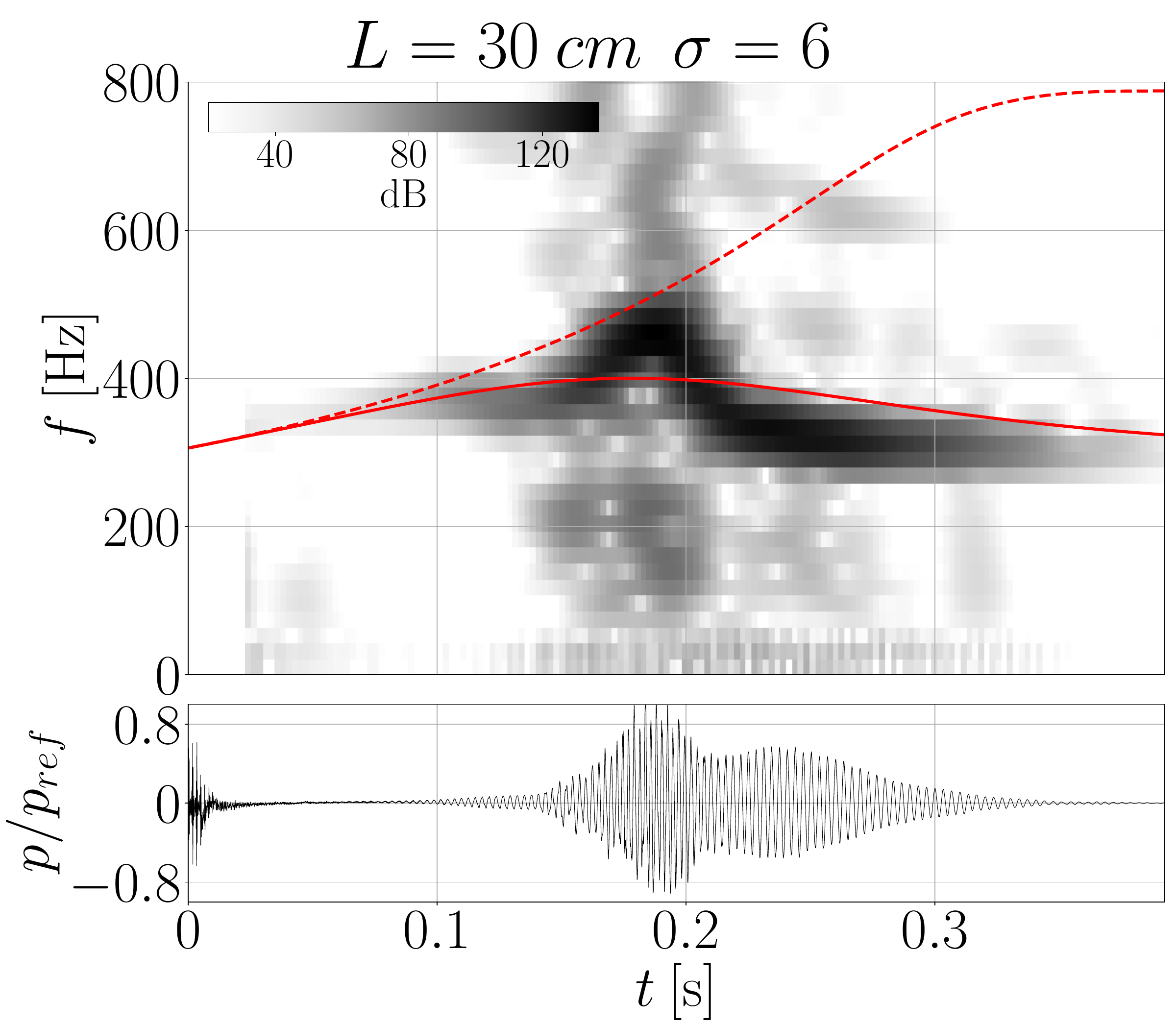}

	\caption{Audio signal and its time-frequency spectrogram for various tube lengths $L$. Unburnt gas temperature is $T_u = 333$~K, temperature ratio $\epsilon = 6.75$. \textcolor{black}{black and blue curves depict the fundamental and first harmonic frequencies respectively, for the adiabatic (dashed) and corrected model (solid).}}
	\label{fig:L100audiospectrogram}
 \end{figure}

A secondary validation of the heat transfer problem can be achieved by systematic modification of the temperature of the working liquid, which sets the wall temperature and the quiescent mixture to a nearly-constant value $T_u$ previous to ignition. Figure~\ref{fig:L100audiospectrogram}, shows the audio signals and their time-frequency transform for a higher temperature of $T_u = 333$~K and tube lengths $L=150$~cm (top left), $L=100$~cm (top right), \textcolor{black}{$L=90$~cm (mid left), $L=70$~cm (mid right), $L=50$~cm (bottom left)} and $L=30$~cm (bottom right). Together with each spectrogram, theoretical dimensional frequency predictions have been plotted for the sake of comparison with \textcolor{black}{$\sigma = 30$, $ 20$, $18$, $ 14$, $10$ and $6$, respectively, with the same value of $l_c =5$~cm}. This comparison requires an additional reconstruction of the temporal domain from the spatial one used in the theoretical prediction. For simplicity, it has been assumed that flames propagate at constant velocity $\overline{u}_f$, averaged in each case to match the approximate duration of the audio signal and the length of the tube. \textcolor{black}{Average flame propagation speed is $\overline{u}_f = 1.3S_L$ for $L = 150$~cm, $ \overline{u}_f =1.2S_L$ for $L =100,\,90$ and $70$~cm, and $\overline{u}_f =1.7S_L$ for $L = 50$ and $30$~cm, with the theoretical planar flame velocity $S_L\simeq0.45\mathrm{m\cdot s^{-1}}$}. However, as it has been mentioned above, front velocity depends strongly on the propagation regime. For instance, flames undergoing primary instabilities see their average speed reduced with respect to the steady propagation velocity. Consequently, horizontal shifting between the model and the experimental results is expected at some sections due to small errors in the temporal reconstruction. The horizontal shifting may become more relevant in shorter tubes $L<60$~cm, due to the onset of violent secondary instabilities and subsequent larger variations of the front velocity. This effect can be inferred from the compressed spectogram in time coordinate in comparison with the theoretical prediction for the $L=30$~cm case. However, good agreement between the model and the experiments is achieved as shown in Fig.~\ref{fig:L100audiospectrogram}, which indicates that the constant velocity assumption renders negligible time shift for most of the cases. 

In addition, it should be noted that the thirty-degree increase of temperature shown in Fig.~\ref{fig:L100audiospectrogram} modifies the transition dynamics to secondary instabilities in the test of $L=100$~cm, as compared to Fig.~\ref{fig:audio_signal_length} results. However, the excited modes are again the first harmonic at the $L_{\rm o.e.}$ and $L_{\rm c.e.}$ regions, and the fundamental mode in the mid-section of the tube. Further comparisons for shorter lengths show the transition from the first harmonic to the fundamental mode in the aforementioned regions, with effective superposition of both frequencies for $L=70$~cm. Moreover, fundamental mode predominance for $L=30$~cm is found, where the correction to the adiabatic case is less pronounced as $l_c\simeq 5$~cm is not so small compared to $L$. Note that the dimensional coupling frequency grows as the tube length is reduced and the acoustic time becomes shorter.

\begin{figure}
    \centering
    \includegraphics[width=1\textwidth]{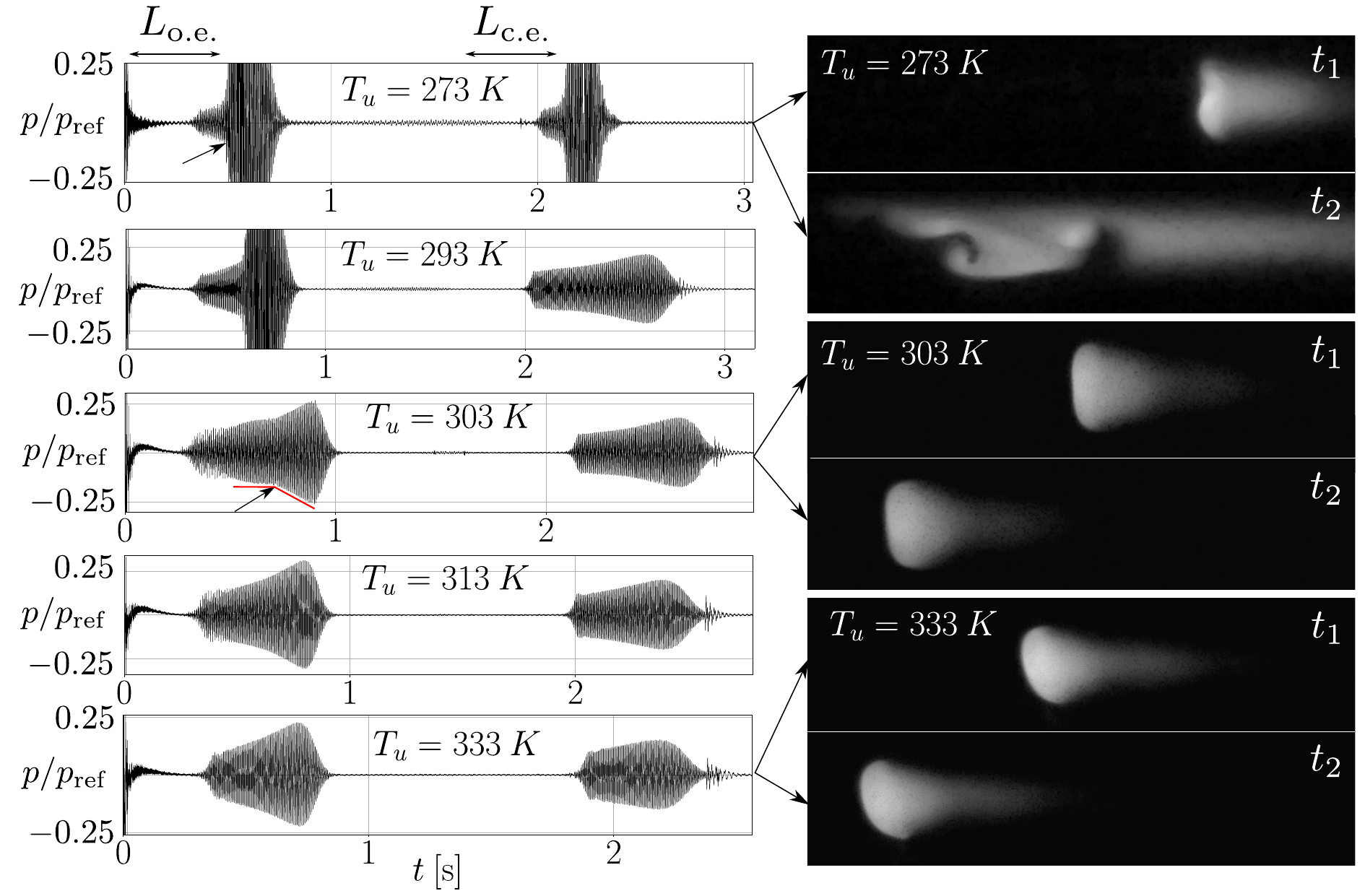}
	\caption{Audio signal and high-speed snapshots taken at the $L_{\rm o.e.}$ region for $L=150$~cm ($\sigma = 30$), $\phi = 1$ and increasing bath temperature $T_u$.}
	\label{fig:Audio_temperature}
 \end{figure}
 
Finally, progressive increase of $T_u$ drives the gradual mitigation of the transition from primary to secondary instabilities as inferred from the audio signals presented in Fig.~\ref{fig:Audio_temperature} for fixed length $L=150$~cm and stoichiometric equivalence ratio $\phi = 1$. The growth in amplitude at the $L_{\rm o.e.}$ coupling region develops an abrupt change in trend at 273~K and again at the $L_{\rm c.e.}$ section, which drive the transition to violent oscillations of the flame front. However, an increase of the bath temperature to 293~K prevents the transition to secondary oscillations at the $L_{\rm c.e.}$ section. Experimental imaging of the cold bath tests indicates that the flame front starts to produce a nearly-sinusoidal front corrugation under the primary weak beating, at time $t_1$ previous to the point marked with an arrow. This perturbation on the reaction surface serves as a precursor of the strong dynamics build up, which later produces a completely corrugated and stretched front as shown in the $t_2$ snapshot of the right panel, after the arrow-indicated instant.

Besides, tests were also performed for a bath temperature of 303~K. This temperature increase results in the suppression of primary-to-secondary transition though the kink in the envelope of the pressure signal is still noticeable. The latter, marked with an arrow, indicates the reminiscence of the transition developed in colder conditions. These details are also explored through the high-speed imaging, where the flame front becomes almost planar with a very subtle curvature that does not evolve to produce a consequential corrugation. In this case, small initial oscillations grow slowly in amplitude without a change in regime. 

Finally, the inflection point that indicates the transition is yet more attenuated into a smooth increase of slope for $T_u=313$~K, and an almost constant growth of amplitude in the pressure signal for $T_u = 333$~K. This fact translates into a nearly-constant flame beating that preserves a more curved front, similar to the steady propagation regime and far from becoming flat nor corrugated. In addition, it must be stated that the cooler cases transitioning to secondary instabilities show a weak arising of the second harmonic in superposition to the first one. However, this effect is weakened for increasing temperatures and the second harmonic influence is strongly reduced.

\section{Discussion}
The experimental work presented above provides detailed insight into the coupling frequency conditions in narrow tubes that are key to the excitation of thermoacoustic instabilities and their transition to large-amplitude high-pressure oscillations. Non-adiabatic considerations have proven useful in the prediction of real frequency coupling of experimental methane flames in slender tubes open at the ignition end.

This has been widely applied to a large variety of lengths and temperature conditions of the thermofluid system with great quantitative agreement. Therefore, the main mechanism driving the first-order frequency coupling is the actual temperature profile of the gaseous domain, hence the local speed of sound distribution controls the thermoacoustics. However, the regime under which these instabilities develop is not just a matter of the excited frequency, but also whether there is transition or not to corrugated strongly-oscillating fronts, which has been shown to be dependent on the temperature prescribed by the thermal bath.

These behaviors are expected to be directly related to three different mechanisms that can influence the dynamics. In the first place, varying temperatures of the wall and mixture produce a modification of the acoustic time through the change in speed of sound, which may be responsible for different frequency coupling. Second, higher bath temperatures promote viscous effects as a dissipative mechanism, which can be key to the damping of higher harmonics with greater oscillation frequencies. At the same time, the viscous layer in a pulsating flow near the walls can be increased to thicknesses that are not negligible compared to the tube radius \citep{veiga2019experimental}. In the third place, the change of conductive heat losses at the wall by modifying its temperature also induce a change in the reaction front curvature at its tip, and the flame border can be placed at various distances of the wall. These effects can be key to the production of ripples at the edge of the flame surface through the pulsatile flow \citep{higuera2019acoustic}, and enable the development of secondary instabilities.

The causes that control this kind of transition are ultimately more difficult to explain in the light of the experiments at hand. Nevertheless, the fuel-to-air ratio for methane-air, \textcolor{black}{propane-air and DME-air} flames was argued to play a major role in controlling such transitions to secondary instabilities in Hele-Shaw cells through the curvature and diffusive effects captured by \textcolor{black}{decreases in the Markstein number as a function of different equivalence ratios of the mixture} \citep{aldredge2004experimental,veiga2019experimental}. Here, that behavior is also recovered, in agreement with a lack of transition for nearly stoichiometric methane mixtures at ambient temperature $T_u\sim 303$~K.

\begin{figure}
    \centering
	\includegraphics[width=0.55\textwidth]{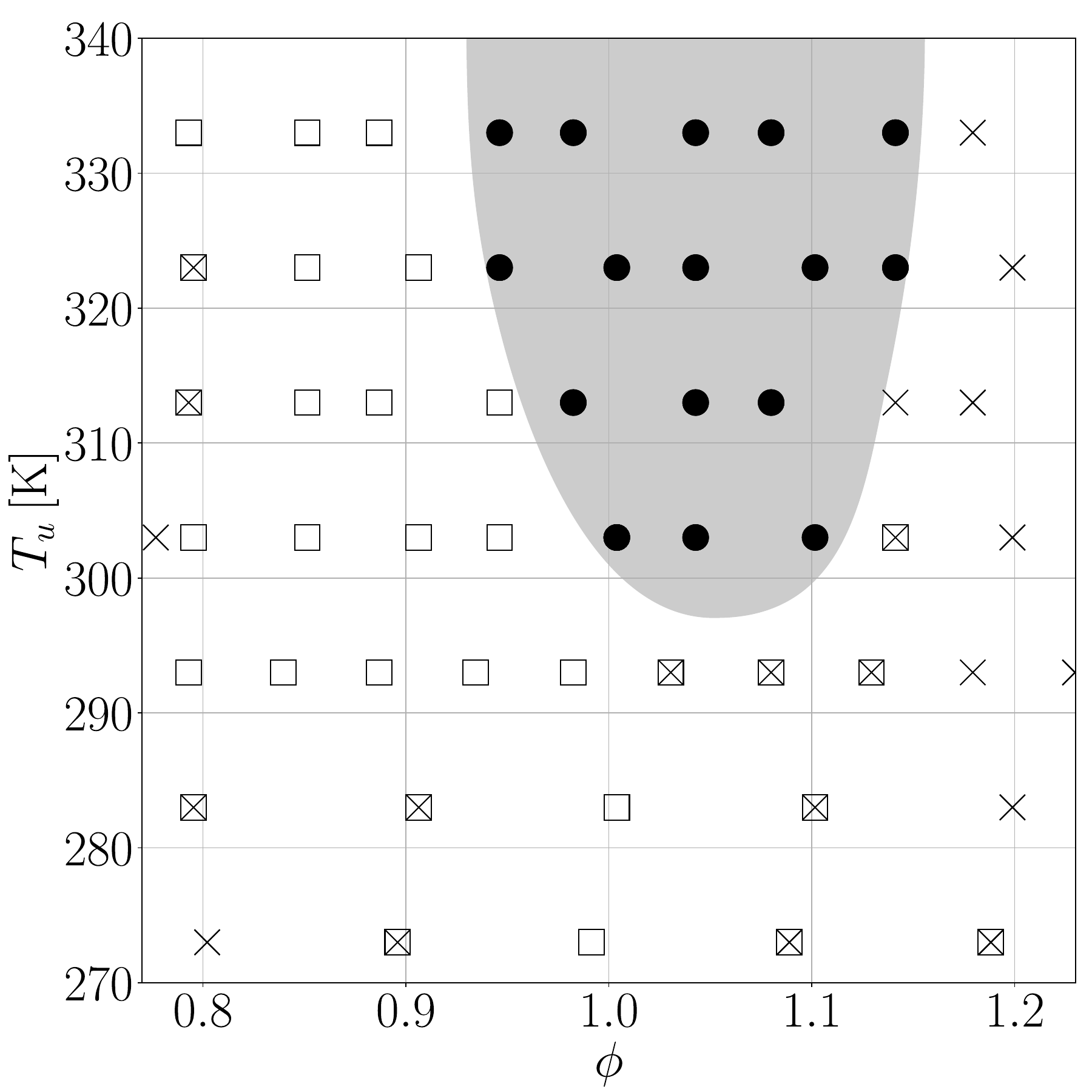}
	\caption{Stability map over the controlling parameters $\phi$ and $T_u$, given by experimental data at the $L_{\rm o.e.}$ region of a $L=150$~cm tube. Primary oscillations of the flame are presented as solid circles, secondary oscillations as empty squares and crosses indicate flame extinction.}
	\label{fig:Parameters_space}
\end{figure}

In addition, it has been shown that in sufficiently long tubes, thermo-acoustic instabilities are constrained to two regions localized around $L_{\rm o.e.}$ and $L_{\rm c.e.}$ sections, measured from the closed end of the tube. Regardless of the equivalence ratio and the unburnt mixture temperature, instabilities always develop at these sections in $L=150$~cm combustion chambers. However, in far-from-stoichiometric mixtures, flames usually extinguish with the onset of secondary instabilities. Since transition to secondary instabilities often occurs first at the $L_{\rm o.e.}$ region, these rich and lean mixtures cause the weaker flame fronts to blow off and the data under study of the subsequent $L_{\rm c.e.}$ region becomes too scarce. Therefore, the influence of $\phi$ and $T_u$ on the instability regimes in the $L_{\rm o.e.}$ region is presented in Figure~\ref{fig:Parameters_space} through a stability map. The shadowed region identified by the experimental results marked with solid circles denote cases for which no transition to secondary instabilities are observed, only developing primary weak oscillations. Moreover, the clear area is conformed by the combination of conditions that produce a transitional behavior to the secondary regime. However, empty squares indicate effective transitions from primary to secondary instabilities and crosses depict transition cases which involve flame extinction due to the violent stretching produced by the flow field. It should be noted that certain combinations of parameters are marked with both symbols, which imply that some experimental runs produce flame extinction and others a successful completion of the transition. Therefore, those cases are in the limiting region between effective progression to secondary instabilities and unbearable flame stretching.

A less unstable region exists at near stoichiometric mixtures and high preignition temperatures. These results emphasize the existence of a stabilizing effect with temperature, as shown by the experimental results of Fig.~\ref{fig:Audio_temperature}. The width of the shadowed region grows with increasing values of $T_u$ and disappears completely for sufficiently cold conditions of the wall-mixture system. For these reasons, further experimental studies need to be carried out in the spirit of providing flow-field measurements and characterization of the wall shear effects in pulsatile flows under varying temperature conditions.

\section{Concluding remarks}

Experimental observations of thermoacoustic coupling in long, $L\simeq 100$~cm, and narrow, $D=10$~mm, tubes with thermal control of wall temperature show that the classic one-dimensional predictions require a theoretical extension to capture non-adiabatic realistic scenarios. First, a theoretical model accounting for heat losses and based on a quasi-one-dimensional heat transfer analysis of the flow downstream of the flame front was provided. This model yields a more complete description of the acoustic field that interacts with the flame in the experiments. The eigenvalues of the non-adiabatic acoustics model have to be computed numerically and depend on both the temperature ratio across the flame $\epsilon$ and on the tube-length-to-cooling-distance ratio $\sigma = L/l_c$. The fixed diameter of the tube and \textcolor{black}{constant temperature of the wall} allow to estimate a constant cooling length $l_c \simeq 5$~cm.

A wide experimental analysis is then performed to explore the effects induced by the variation of the controlling parameters of tube length and wall temperature via an adjustable setup. In particular, a movable piston and a circulating thermal bath are used to improve the experimental results. Now, measured frequency values are in much better agreement with the proposed non-adiabatic model. The results presented above bring to light how heat losses affect the acoustic frequency via modification of the burnt gases temperature and therefore the natural acoustic frequencies of the tube. Specifically, test on long tubes ($L\simeq150$~cm) show a development of two main regions near the ends of the tube where the first harmonic arises in the coupling thermoacoustic dynamics. Shorter tubes in the intermediate range ($L\simeq 90$~cm) promote the excitation of the fundamental acoustic mode in the mid region of the chamber in addition to the previous coupling. Finally, this fundamental mode is mainly preserved for the shortest cases ($L\simeq 30$~cm) in absence of the first-harmonic coupling near the ends.

In addition, the reacting front area has been explored in relation with the transition to secondary thermoacoustic instabilities. It has been found that progressive heating of the mixture and walls stabilize the flame, preventing the surface from rippling and developing disordered strong oscillations. Therefore, long thin tubes under conductive heat-loss effects and increasing viscous dissipation open the way to primary thermoacoustic instabilities only. This particularity is here reported for the first time in combination with an equivalence ratio variation to produce a thermoacoustic stability map. More stable scenarios have been found for nearly-stoichiometric methane-air mixtures and bath temperatures above $300$~K.

 \section*{Acknowledgements}
\noindent

The authors would like to acknowledge the technical support of Fernando Fernández Arcas and Jos\'e Miguel Velasco Gómez in the construction of the experimental setup. 

\noindent
\textbf{Funding}{ This work was funded by the Agencia Estatal de Investigación of Spain (PID2019-108592RA-C43 / AEI / 10.13039/501100011033) \& (PID2019-108592RB-C41 / AEI / 10.13039/501100011033) }

\noindent
\textbf{Declaration of interests.}{ The authors report no conflict of interest.}

\noindent
\textbf{Author ORCIDs}{\\
E. Flores-Montoya, https://orcid.org/0000-0002-1783-7507;\\
V. Muntean, https://orcid.org/0000-0001-8096-1773;\\
M. S\'anchez-Sanz, https://orcid.org/0000-0002-3183-9920;\\ 
D. Martínez-Ruiz, https://orcid.org/0000-0001-6233-213X.}

\appendix

\section{Data Processing }\label{appendix:postprocess}

High-speed videos register the luminous emission of the flame as it propagates along the tube. It is to be pointed out that the camera is cautiously positioned using optical alignment tooling, so that the plane of the sensor is kept parallel to the longitudinal axis of the tube. Images can be interpreted as a function of space that takes $8$-bits integer values and is sampled in time by the camera's frame rate, $f(x_m,y_n,t_i)$, where $m$ and $n$ are the number of columns and of rows of the image respectively, and $i$ is the frame number.

In-house post-processing software computes the relative displacement in the streamwise direction $x$ of the local flame front, $\Delta x(y_n, t_i)$, between frames $i$ and $i+1$ for each point $y_n$ using correlation analysis. Later on, the flame velocity $u(y_n,t_i)$ and position $x(y_n,t_i)$ can be computed from the knowledge of this displacement and the time lapse, $\Delta t = t_{i+1}-t_i$. The time lapse is the same for all pairs of consecutive frames. For the calculation of $\Delta x(y_n, t_i)$ first the discrete cross-correlation between two frames is computed,
\begin{equation}
   R_{xx}^i(\zeta_n, y_n) = \sum_{k=0}^m f_x(x_k,y_n, t_i)f_x(x_k+ \zeta_n, y_n,t_{i+1}),
\end{equation}
where subindex $x$ denotes the discrete longitudinal derivative of function $f(x_m,y_n,t_i)$ and $\zeta_n$ is a discrete variable. The use of $f_x$ instead of $f$ prevents the flame tail from introducing a bias in the correlation maximum and therefore a drift in the flame position tracking. The value of $\zeta_n$ that maximizes the correlation function $R_{xx}^i(\zeta_n, y_n)$, represents the local displacement of the flame front $\Delta x(y_n, t_i)$, and it is different for each row of the image. A subpixel interpolation method is employed to increase the accuracy of the calculation of the displacement. The latter consists in approximating the vicinity of the maximum of the correlation function by a parabola for each $y_n$. The position of the maximum of the obtained parabola allows the computation of non-discrete values for the displacement. Then, the average displacement of the flame front is computed as $\Delta \bar{x}(t_i) = \sum_n \Delta x( y_n, t_i)/n$ and the instantaneous velocity of the flame front is calculated as $u(t_i) = \Delta \bar{x}(t_i)/\Delta t$. Fourier analysis is performed over the velocity signal $u(t_i)$ to obtain the frequency components of flame movement.

\subsection{Uncertainty evaluation}
\textcolor{black}{
In order to estimate the error involved in the determination of the presented experimental data, an uncertainty error analysis was performed. The error in the determination of the dimensionless frequency, $ \Omega = \omega L/c_u $, comes from the finite frequency resolution of the Fourier analysis, the uncertainty in adjusting the tube's length, $L$, and the uncertainty of the water flow temperature, $T_u$, that affects the calculated value of $c_u$. Applying the method of propagation of uncertainty, see \cite{moffat1988describing}, the sought error is given by
\begin{equation}
 \Delta \Omega = \sqrt{\left(\dfrac{\partial\Omega}{\partial \omega}\Delta\omega\right)^2 + \left(\dfrac{\partial\Omega}{\partial L}\Delta L\right)^2+ \left(\dfrac{\partial\Omega}{\partial c_u}\dfrac{\partial c_u}{\partial T_u}\Delta T_u\right)^2}.
 \label{eq:uncert}
\end{equation}
In this expression  the  uncertainty of the signal's frequency is $\Delta \omega = 2\pi f_s /n$, where $f_s$ is the acquisition frequency that matches the frame rate of the high speed camera and $n$ is the number of samples of $u(t_i)$ used in the Fourier analysis.
}
\textcolor{black}{
During the experiments the position of the piston was manually driven and directly affected the accuracy of $L$. The position of the piston was controlled by means of several marks on it with a separation of $1~\mathrm{cm}$ between them. This allows us to ensure that the absolute error in the piston's placement is $\Delta L = 5~\mathrm{mm}$. With regard to the error in $T_u$ measurements, the worst case scenario is supposed and the typical uncertainty of a thermocouple is used, $\Delta T_u = 1~\mathrm{K}$.
}
\textcolor{black}{
The uncertainty in the value of the dimensionless frequency $\Omega$ is determined using (\ref{eq:uncert}), and corresponds to the vertical error bars in Fig.~\ref{fig:video_frequencies}. However, it is worth mentioning that the main contribution to the uncertainty propagation is due to the Fourier analysis of the velocity signal. The evaluation of the uncertainty for a typical case with $L =0.7$~m, $T_u  = 303$~K, $f_s = 2000$~Hz , $f = 400$~Hz and $n = 150$ yields
\begin{equation}
    \Delta \Omega = \sqrt{\left(0.168\right)^2 + \left(0.036\right)^2+\left(0.008\right)^2} \simeq 0.172,
\end{equation}
where it is noteworthy that the errors introduced by the uncertainty in the combustion chamber's length and in the ambient temperature are much smaller than the error due to the limited resolution of the Fourier analysis. The resulting relative error turns out to be $\Delta \Omega/\Omega\simeq 3.4\%$ in that case. In the experimental results of Fig.~\ref{fig:video_frequencies}, relative errors range between $\Delta \Omega/\Omega\simeq 2.9\%$ and $\Delta \Omega/\Omega\simeq 14.3\%$ depending on the tube's length and the number of samples $n$.
}

\begin{figure}
    \centering
	\includegraphics[width=1\textwidth]{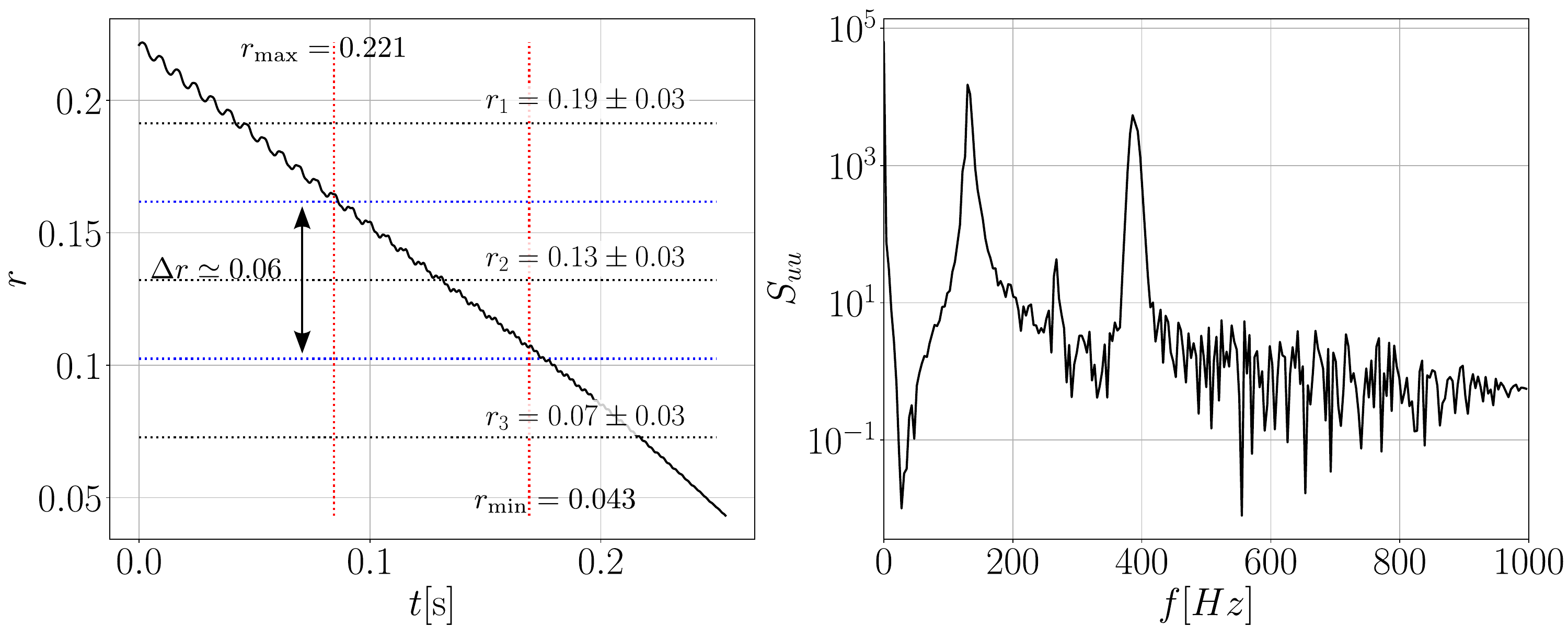}
	\includegraphics[width=1\textwidth]{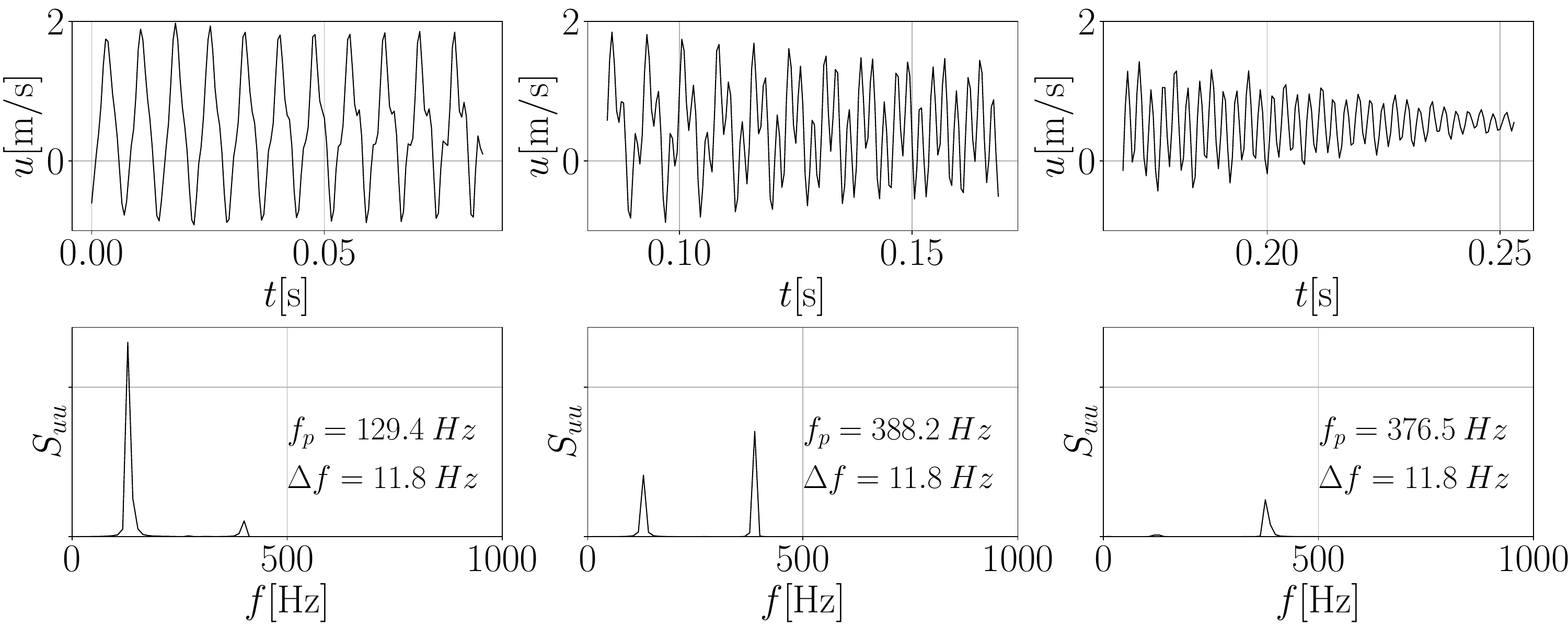}
	\caption{\textcolor{black}{Uncertainties and Fourier Analysis for a particular experimental run.}}
	\label{fig:Errors_Fourier_analysis}
\end{figure}
\textcolor{black}{
It is also worth to clarify here the uncertainty in the position of flame, $\Delta r$. The discussion will be presented for a particular case, but it can be generalized for all the experimental runs.
The  dimensionless position of the flame obtained from the processing of the recorded images for a particular experimental run is shown in the top left panel of Fig.~\ref{fig:Errors_Fourier_analysis}. As it can be noted, in this run $r$ goes to zero, so the flame is reaching the end of the tube. The presented values of $r$ are obtained by knowing the placement of the camera with respect to the tube and the scale in mm/pixel. The portion of the tube in the field of view of the camera is $\hat{r}\in[r_{min},\>r_{max}]$, but in order to increase the resolution of the experimental data this interval is split into $m$ parts of equal duration,
\begin{equation}
    \hat{r}_k = \{[r_{min},\>r_1], \>[r_1,\>r_2],\ldots,[r_{m-1}, r_{max}]\},
\end{equation}
with $k=1,2\ldots,m$.
}
\textcolor{black}{
The velocity signal, $u(t_i)$, is also split into as many equal length parts as the position signal, the resulting signals being $u_k(t_i)$ with $k=1,2\ldots,m$. The spectrum, $S_{u_ku_k} = \tilde{u_k}\circ\overline{\tilde{u_k}}$ with $\tilde{u_k}$ its Fourier transform and $\overline{\hat{u_k}}$ its complex conjugate, of each part of the velocity signal is computed and the frequency $f_{pk}$ of the peak is identified. As the uncertainty in the determination of the frequency is given by $\Delta f = mf_s/N$, it is important to notice that the value of $m$ is the result of a compromise between the resolution and the uncertainty in the experimental results. For the experimental run presented in Fig.~\ref{fig:Errors_Fourier_analysis}, the value of $m$ is $3$ and  $\Delta f = 11.8$~Hz.
}

\textcolor{black}{
Finally, the  black dotted lines in the top left panel of Fig.~\ref{fig:Errors_Fourier_analysis} mark the dimensionless flame position $r$ that is assigned to the frequency of each interval $k$. The frequency $f_{pk}$ of the maximum of the spectrum $S_{u_ku_k}$ for the velocity signal $u_k(t_i)$ is assigned to the position $(r_k + r_{i-k})/2$ and the spatial uncertainty is taken as half the length of each of those intervals $\Delta r = (r_k - r_{k-1})/2$. Thus, for the presented experimental run, the frequency is determined for three different positions of the flame: $r=0.07,\ 0.13\ \mathrm{and}\ 0.19$ with an uncertainty of $\Delta r = 0.03$ each, directly related to the horizontal error bars of Fig.~\ref{fig:video_frequencies}. The values of the frequencies for these positions of the flame are $f_p=129.4,\ 388.2\ \mathrm{and}\ 376.5$~Hz respectively with an uncertainty of $\Delta f = 11.8$~Hz each as it can be observed in the bottom line of graphs in Fig.~\ref{fig:Errors_Fourier_analysis}. The calculated frequencies are then nondimensionalized and the obtained values of $\Omega$ are plotted in Fig.~\ref{fig:video_frequencies}.
}
\subsection{Repeatability}
\begin{figure}
    \centering
	\includegraphics[width=1\textwidth]{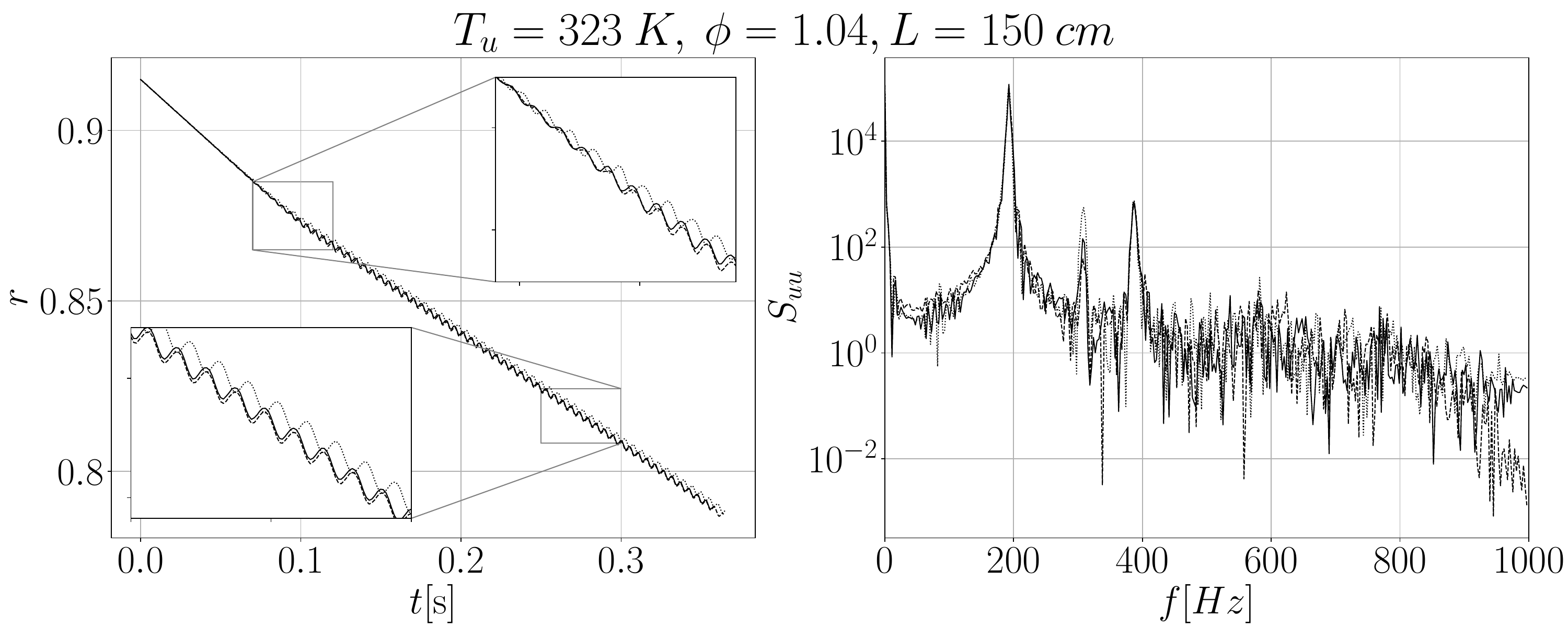}
	\includegraphics[width=1\textwidth]{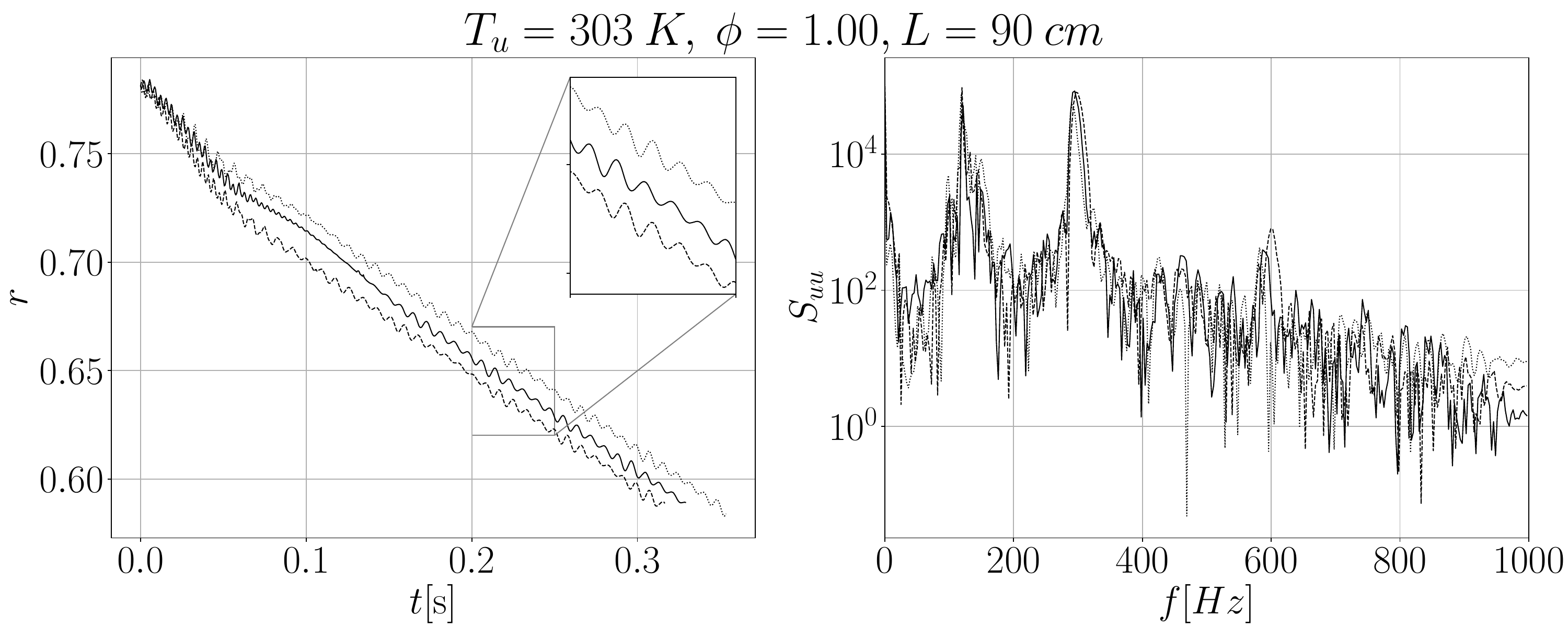}
	\caption{\textcolor{black}{Experimental repeatability, flame position as obtained from the high-speed video and and spectrum of velocity signal in three different runs (solid, dashed, dotted).}}
	\label{fig:repeatability}
\end{figure}

\textcolor{black}{Regarding repeatability, every parametric combination of $\phi$, $T_u$ and $L$ was tested at least three times. The signals of flame position as extracted from high-speed imaging and the spectrum of the velocity signals are presented in Fig.~\ref{fig:repeatability} for two cases. Solid, dashed and dotted lines depict the different runs under the conditions stated in the figure. Highly-repeatable data is obtained from a visual inspection of the flame position as from the Fourier frequency analysis. As it can be observed, the position of the flame is not exactly the same for different experimental runs with the same configuration of controlling parameters. The position is shifted depending on the initial transition from an uniform to a vibrational mode of propagation of the flame front. The result of this behaviour is that the flame front propagates similarly in different runs, but the evolution may be slightly-shifted in time. When the dominant frequencies are inspected in the spectrum of the velocity signal, it is observable that in spite of the shifting the repeatability is adequate.}

\bibliographystyle{jfm}
\bibliography{paperbib_tuneao}

\end{document}